\documentclass[12pt,letterpaper]{article}

\usepackage{amsmath,amssymb} 

\usepackage{geometry}
\usepackage{graphicx}
\usepackage{xcolor}

\usepackage{float} 
\usepackage[section]{placeins}

\usepackage[labelfont={bf,small},textfont={small},format=hang]{caption}

\usepackage{subcaption}
\usepackage{array}
\usepackage{longtable}
\usepackage{booktabs} 
\usepackage{threeparttable}
\usepackage{multirow}
\usepackage{placeins}
\usepackage{rotating}
\usepackage{siunitx}   

\usepackage{enumitem}

\usepackage{authblk}

\usepackage[round,authoryear,longnamesfirst]{natbib}
\usepackage[colorlinks=true,
            linkcolor=black,
            citecolor=black,
            urlcolor=black]{hyperref}
\usepackage{cleveref}

\usepackage{setspace}
\makeatletter
\renewcommand\@seccntformat[1]{\csname the#1\endcsname.\quad}
\makeatother

\renewcommand{\thesection}{\Roman{section}}          

\usepackage{footmisc}

\title{Squeezed Covariance Matrix Estimation:\\Analytic Eigenvalue Control}

\author{Layla Abu Khalaf}
\author{William S.~Smyth\thanks{Email: w.smyth@ulster.ac.uk}}

\affil{Intelligent Systems Research Centre, Ulster University, Derry, UK}

\date{\today} 

\begin{document}
\maketitle


\begin{abstract}
We revisit Gerber’s Informational Quality (IQ) framework, a data-driven approach for constructing correlation matrices from co-movement evidence, and address two obstacles that limit its use in portfolio optimization: guaranteeing positive semidefiniteness (PSD) and controlling spectral conditioning. We introduce a squeezing identity that represents IQ estimators as a convex-like combination of structured channel matrices, and propose an atomic-IQ parameterization in which each channel-class matrix is built from PSD atoms with a single class-level normalization. This yields constructive PSD guarantees over an explicit feasibility region, avoiding reliance on ex-post projection. To regulate conditioning, we develop an analytic eigenfloor that targets either a minimum eigenvalue or a desired condition number and, when necessary, repairs PSD violations in closed form while remaining compatible with the squeezing identity. In long-only tangency backtests with transaction costs, atomic-IQ improves out-of-sample Sharpe ratios and delivers a more stable risk profile relative to a broad set of standard covariance estimators.
\end{abstract}

\section{Introduction}

Covariance and correlation matrices are central to portfolio construction, risk measurement, and pricing, yet high-dimensional estimates are noisy and often ill-conditioned. Classical responses such as linear and nonlinear shrinkage \citep{ledoit2004honey,ledoit2004well,ledoit2017,ledoit2022quadratic}, random-matrix cleaning \citep{laloux1999,bun2017cleaning}, factor models \citep{chamberlain1983factor,fan2013factor}, and ex-post PSD projection \citep{higham2002computing} stabilize the sample estimator or its spectrum. These approaches improve behavior in practice, but none provide analytic control of eigenvalues: shrinkage intensities are estimated asymptotically, RMT rules are approximate, and Higham’s projection enforces PSD iteratively. No existing method offers closed-form eigenvalue targets for a given estimator.

Gerber Informational Quality (IQ) took a different route: it builds correlation matrices directly from structured co-movement statistics, producing interpretable estimates tailored to task objectives. While IQ often improved conditioning relative to the sample covariance, it did not guarantee PSD and, like other methods, offered no explicit means of controlling eigenvalues.

We develop \emph{atomic--IQ}, a constructive refinement of IQ that addresses both limitations. It introduces the canonical squeezing identity, which can be viewed as balancing a baseline prior of mutual independence (a neutral benchmark) with structured dependence extracted from co-movement evidence. Each class matrix is built from positive semi-definite atoms and normalized once at class level, yielding correlation-PSD channels; under controllable conditions the full estimator is PSD without ex-post repair. We characterize feasibility through an exact spectral condition and give closed-form bounds in the basic--IQ case (squeezing channel weights $\{\eta^2,\eta,1\}$ with $\eta\in[0,1]$). Finally, we introduce an eigenfloor which not only raises $\lambda_{\min}$, contracts $\lambda_{\max}$, and provides closed-form rules for targeting a floor or condition number, but also acts as an analytic PSD repair that remains inside the squeezing representation. This yields the first covariance estimation framework with closed-form eigenvalue control. Implementation details and practical guardrails for using atomic–IQ in risk systems, together with additional spectral results and the full set of Sharpe ratio tests relative to atomic squeezing, are presented in the appendix.

The broader literature on covariance estimation can be divided into two families. The first consists of sample-anchored regularizers. Linear and quadratic shrinkage blend the sample covariance with a structured target, while nonlinear shrinkage modifies eigenvalues to reduce sampling error \citep{ledoit2004well,ledoit2017,ledoit2022quadratic}. Random-matrix theory (RMT) methods remove or adjust modes in the Mar\v{c}enko--Pastur bulk \citep{laloux1999,bun2017cleaning}. Factor models \citep{chamberlain1983factor,fan2013factor} replace the full system with a low-rank latent representation. Higham’s algorithm repairs indefiniteness ex post \citep{higham2002computing}. Despite their differences, all of these methods begin with the empirical covariance (or correlation) matrix and then regularize it after the fact. 

The second family consists of concordance-based constructive estimators. These build correlation estimates directly from co-movement or concordance statistics rather than from the sample covariance. Kendall’s $\tau$, Spearman’s $\rho$, and the Gerber statistic exemplify this approach, as does IQ. Such methods are constructive in that they assemble correlation matrices from structured evidence, including ranks, signs, or thresholded events, thereby bypassing the sample covariance altogether.

IQ belongs to this constructive family but also extends it. Whereas earlier concordance methods provide pairwise measures, IQ generalizes the approach to a system-wide framework. Its $\delta$--$\eta$ template aggregates concordant and discordant events into interpretable class matrices in a way that is compatible with modern optimization and machine learning. Atomic--IQ strengthens this framework by ensuring PSD through atomic construction and by introducing explicit eigenvalue controls via the eigenfloor. In this way IQ not only broadens the constructive family but also provides the first interpretable covariance estimator with analytic eigenvalue control. With atomic--IQ, the earlier concern that IQ matrices might fail to be PSD is resolved, and the approach can be regarded on the same footing as conventional methods while retaining its distinctive constructive character.

In financial applications this matters directly: portfolio optimization, risk-parity, and risk-management objectives are highly sensitive to the spectrum of the covariance matrix. Analytic eigenvalue control within the squeezing framework therefore provides not only structural validity but also transparent and tunable stability, allowing covariance estimates to be aligned explicitly with optimization requirements.

\section{The Gerber Informational Quality Framework}

The Gerber-IQ framework was designed to construct correlation matrices directly from structured co-movement statistics, bypassing the sample covariance and its associated noise. At its core lies a squeezing template, which maps pairs of asset returns into an alignment space and applies structured thresholds to separate informative co-movements from noise. The framework is parameterized by a collection of functional parameters that govern how evidence is aggregated:  
\begin{itemize}
    \item $\mathbf{c}=\mathbf{c}(r_0, c)$, which aligns marginal distributions on $r_0$ and defines the exclusion region for noise through $c$,  
    \item $\boldsymbol\delta$, which sets the boundaries of the squeezing channels,  
    \item $\boldsymbol\eta$, which assigns squeezing weights to the various channels or channel classes,  
    \item $\gamma$, which governs temporal squeezing by reflecting the predictive value of signal based on recency,  
    \item $\epsilon$, which acts as a delay parameter for $\gamma$-activation, and  
    \item $\tau$, which specifies the lookback duration for the estimation sample.  
\end{itemize}

We sometimes represent the collection of spatial parameters as $\mathbf{s}=\mathbf{s}(\mathbf{c},\boldsymbol{\delta},\boldsymbol{\eta})$ and the temporal parameters as $\mathbf{t}=\mathbf{t}(\tau,\epsilon,\gamma)$.
These parameters can be set by an experienced analyst or, in part or in whole, learned within an optimization framework, for example by deep learning architectures such as Markowitz-Informed Neural Networks \citep{minns2025ssrn}.  

\begin{figure}[ht]
    \centering
    \includegraphics[width=0.9\textwidth]{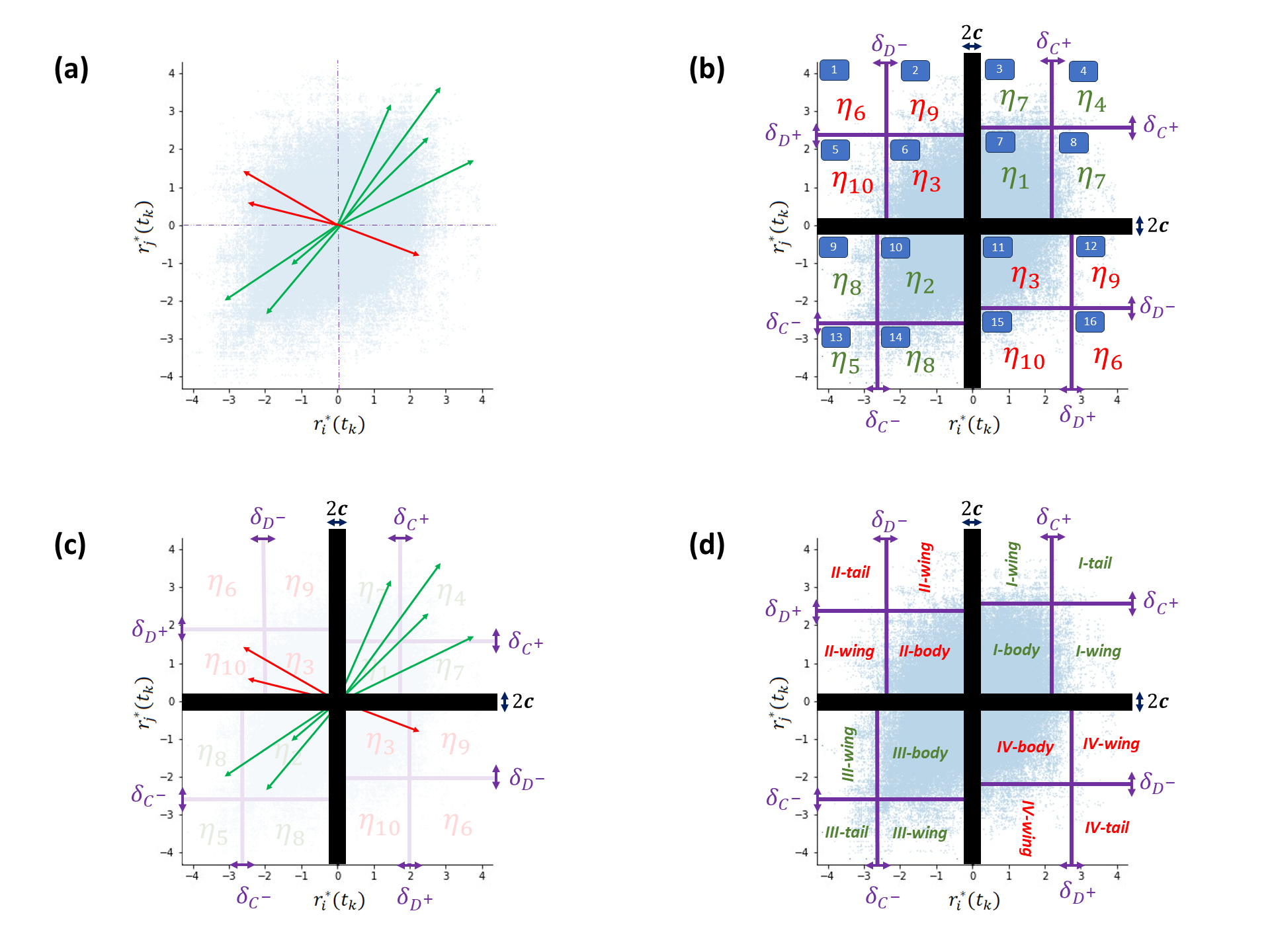}
    \caption{
        Illustration of the $\boldsymbol{\delta}$–$\boldsymbol{\eta}$ alignment template: 
        co-movement vectors (a), squeezing channels (b), vectors on the template (c), and 
        refined body–tail–wing structure (d).
    }
    \label{fig:fig_1}
\end{figure}

Figure~\ref{fig:fig_1} illustrates this alignment structure.  
Panel (a) shows sample co-movement vectors; Panel (b) depicts the squeezing channels defined by $\boldsymbol{\delta}$ and $\boldsymbol{\eta}$; Panel (c) overlays co-movement vectors 
onto the template; and Panel (d) refines the structure into body, wing, and tail regions (or channel classes). This representation provides a clear and interpretable map from raw return pairs to a 
structured statistical template. 

To compute IQ we refer to the squeezing template in Figure~\ref{fig:fig_1}. This structure translates into statistical expressions that generate correlation matrix elements $\rho_{ij}(t; \mathbf{s},\mathbf{t})$, representing the IQ measure of co-movement between assets $R_i$ and $R_j$ at time $t$. Let $T:=\{0,1,\dots,\tau-1\}$ index the $\tau$ observations in the lookback window, ordered from oldest to most recent, with times $\{t_m\}_{m\in T}$. For each asset $R_k$, we define an exclusion region for transformed returns based on the noise threshold $c$:  

\[
M_k := \{ m \in T : |\tilde r_k(t_m)| \leq c \}.
\]

The threshold $c$ may be specified in asset-specific units, for example as a multiple of the sample standard deviation of asset $k$, or in pairwise units based on an aggregate of the volatilities of assets $i$ and $j$ (such as $\min\{\hat\sigma_i,\hat\sigma_j\}$, $\max\{\hat\sigma_i,\hat\sigma_j\}$, or an average). From these, we form index sets over which the statistic is computed:  

\[
E_\cup := T \setminus (M_i \cup M_j), \qquad
E_\cap := T \setminus (M_i \cap M_j).
\]

We then define indicator functions for concordant and discordant co-movement:  

\[
I^+(t_m) =
\begin{cases}
1 & \text{if } \tilde r_i(t_m)\,\tilde r_j(t_m) > 0, \\
0 & \text{otherwise},
\end{cases}
\qquad
I^-(t_m) =
\begin{cases}
1 & \text{if } \tilde r_i(t_m)\,\tilde r_j(t_m) < 0, \\
0 & \text{otherwise}.
\end{cases}
\]

Let $\eta(t_m; \omega(\mathbf{c}, \boldsymbol{\delta}))$ denote the squeezing weight assigned to the co-movement $(\tilde r_i(t_m), \tilde r_j(t_m))$, where the channel $\omega(\mathbf{c}, \boldsymbol{\delta})$ is determined by the spatial parameters $\mathbf{c}$ and $\boldsymbol{\delta}$. The IQ squeezing statistic is then defined as  

\begin{equation}
\rho_{ij}(t; \mathbf{s},\mathbf{t}) \;=\;
\frac{
\sum_{m \in E_\cup} \Delta(t_m)\, \eta(t_m; \omega(\mathbf{c}, \boldsymbol{\delta}))\, v(t_m;\mathbf{t})
}{
\sum_{m \in E_\cap} \eta(t_m; \omega(\mathbf{c}, \boldsymbol{\delta}))\, v(t_m;\mathbf{t})
},
\label{eq:rho-ij}
\end{equation}

where $\Delta(t_m) = I^+(t_m) - I^-(t_m)$. The numerator aggregates evidence of concordant and discordant co-movements when both transformed returns exceed the noise threshold $c$, while the denominator aggregates evidence when at least one transformed return exceeds the threshold.  

Temporal effects are incorporated through the discount factor  

\[v(t_m;\mathbf{t}) = \exp\!\Big(-\,\gamma\,(\tau-1-(m+\epsilon))_+\Big),
\qquad m\in\{0,1,\dots,\tau-1\}.\]

where $\epsilon$ is the delay parameter, $\gamma>0$ is the decay parameter, and $\tau$ denotes the lookback window duration, which determines the index set over which $v(t_m;\mathbf{t})$ is evaluated.

This element-level construction translates naturally to the system-wide representation. At the matrix level, IQ correlation matrices can be expressed through the canonical squeezing identity:

\begin{equation}
S \;\equiv\; \sum_{\alpha \in \mathcal{K}} \eta_\alpha \, C^{(\alpha)}
\;+\;
\left( 1 - \sum_{\alpha \in \mathcal{K}} \eta_\alpha \right) I,
\label{eq:canonical-squeezing}
\end{equation}

where the coefficients $\{\eta_\alpha\}$ are channel-class squeezing weights and each $C^{(\alpha)}$ is a channel-class matrix. Channels are the non-overlapping regions of the bivariate support defined by the template parameters $\boldsymbol{\delta}$ and $\mathbf{c}$. In practice, channels are grouped into broader channel classes, $\alpha \in \mathcal{K}$, such as body, wing, and tail (Figure 1(d)), with all channels in a class allocated the same squeezing weight. The matrices $C^{(\alpha)}$ should therefore be interpreted as correlation matrices constructed from data passing through particular channel classes. Operational details for implementing the body, wing, and tail channel classes, and the noise-exclusion band, are provided in the appendix.
 
In the next section we turn to how these channel-class matrices are constructed from atomic building blocks and why this guarantees positive semi-definiteness by design.

\section{Atomic--IQ: PSD by Design}

The canonical squeezing identity expresses the estimator as a balance between a zero‑correlation benchmark and a structured component assembled from co‑movement evidence. In this view, \eqref{eq:canonical-squeezing} implements an endogenous zero‑correlation prior, while the class matrices $\{C^{(\alpha)}\}$ provide data‑driven, objective-led adjustments. The coefficients $\{\eta_\alpha\}$ tune the trade‑off, allowing the estimator to range from near‑neutral (close to zero correlation) to strongly structured dependence. The neutral share $1-\sum_\alpha \eta_\alpha$ keeps a measurable portion of the estimate anchored at the benchmark, and the structured share $\sum_\alpha \eta_\alpha C^{(\alpha)}$ moves the estimate away from it in directions and magnitudes supported by the evidence. In short, neutrality is not discarded; it is modulated according to where the information lies across statistical and temporal channels.

An immediate qualitative feature follows. The collective squeeze $\sum_\alpha \eta_\alpha$ determines the overall balance between the neutral benchmark and the structured component. When the collective squeeze is less than one, PSD is guaranteed. Once it exceeds one, PSD is no longer automatic and holds only under additional spectral conditions. The remainder of this section turns to how atomic–IQ constructs the structured component from positive semi‑definite building blocks, and why PSD is preserved when the collective squeeze does not exceed unity.

\subsection{Building Blocks: Atoms, Aggregation and Scaling}

At the event level each concordant/discordant co‑movement between assets $R_i$ and $R_j$ contributes a $2\times2$ atom
\begin{equation}
A^{+}=\begin{bmatrix}1&1\\[2pt]1&1\end{bmatrix},\qquad
A^{-}=\begin{bmatrix}1&-1\\[2pt]-1&1\end{bmatrix},
\label{eq:atoms}
\end{equation}
which is embedded into the $(i,j)$ block of the $n \times n$ correlation template via the selector
\[
J_{ij} := [\,e_i\ e_j\,]\in\mathbb{R}^{n\times 2},\qquad X \mapsto J_{ij} X J_{ij}^\top.
\]
Let $e=(i,j,t)$ index an event involving $R_i$ and $R_j$ at time $t$, with sign $\pi_e\in\{+,-\}$. We may represent a per‑event scaled atom as
\begin{equation}
{A}_e := \mathcal{S}_e\!\left[A^{\,\pi_e}\right],
\label{eq:scaled-atom}
\end{equation}
where $\mathcal{S}_e$ is a PSD‑preserving scaling operator. Two concrete choices (whole‑atom vs off‑diagonal‑only) are given later; for now \eqref{eq:scaled-atom} covers both and implicitly carries any temporal or magnitude effects. Events are grouped into classes $\alpha \in \mathcal{K}$ by the squeezing template. The class accumulator matrix is the sum of embedded scaled atoms:
\begin{equation}
G_\alpha = \sum_{e\in\alpha} J_{ij}\,{A}_e\,J^{\phantom{\top}}_{ij}{}^\top.
\label{eq:Galpha-general}
\end{equation}
If $\mathcal{S}_e$ preserves positive semi‑definiteness, each addend in \eqref{eq:Galpha-general} is PSD and hence $G_\alpha\succeq0$. When all qualifying events in a lookback window have been accounted for, each class aggregator $G_\alpha$ is normalized to correlation scale,
\begin{equation}
C^{(\alpha)} = D_\alpha^{-\tfrac12}\,G_\alpha\,D_\alpha^{-\tfrac12},
\qquad D_\alpha:=\mathrm{diag}(G_\alpha),
\label{eq:class-normalisation}
\end{equation}
which preserves PSD status. Thus, each class matrix $C^{(\alpha)}$ is a PSD correlation matrix. A useful by‑product is refinement invariance: splitting a class into sub‑channels, summing first, and normalizing once yields the same $C^{(\alpha)}$ as treating the class as a single block.

\medskip\noindent
Eqs.~\eqref{eq:Galpha-general}–\eqref{eq:class-normalisation} complete the structured component; the squeezing identity then blends the class matrices with the neutral benchmark as in Section~4.

There are two scaling modes for $\mathcal{S}_e$: whole‑atom scaling (atomic-IQ1) and off-diagonal scaling (atomic-IQ2). In whole-atom scaling the entire $2\times2$ atom receives the same positive weight $a_t$:
\begin{equation}
\mathcal{S}_e^{\mathrm{IQ1}}\!\left[A^{\,\pi_e}\right]
= v_t\,A^{\,\pi_e}
= \big(\sqrt{a_t}\,I_2\big)\,A^{\,\pi_e}\,\big(\sqrt{a_t}\,I_2\big),
\label{eq:IQ1}
\end{equation}

a PSD‑preserving congruence. This modulates qualifying movement and co‑movement together. In off‑diagonal scaling, qualifying movement (diagonal) and co‑movement (off‑diagonal) are decoupled by applying a PSD mask entrywise (Hadamard product). Let

\begin{equation}
H(v)=
\begin{bmatrix}
1 & v\\[2pt]
v & 1
\end{bmatrix},\qquad |v|\le 1,
\label{eq:H-mask}
\end{equation}
so $H(v)\succeq 0$. For event $e$ with sign $\pi_e$, we define
\begin{equation}
\mathcal{S}_e^{\mathrm{IQ2}}\!\left[A^{\,\pi_e}\right]
= H(v_t)\circ A^{\,\pi_e}
= \tfrac{1+v_t}{2}\,A^{\,\pi_e}+\tfrac{1-v_t}{2}\,A^{-\,\pi_e}.
\label{eq:IQ2}
\end{equation}
Since both $H(v_t)$ and $A^{\,\pi_e}$ are PSD, the Schur product theorem ensures $\mathcal{S}_e^{\mathrm{IQ2}}\!\left[A^{\,\pi_e}\right]\succeq 0$. In effect, events retain their occurrence counts, while their co-movement is scaled by $v_t$. 

The profiling factor $v_t$ is determined through temporal scaling. Patterns of co-movement change over time: regimes shift, volatility clusters, and the strength of association between assets may vary within a given lookback window. Temporal scaling lets the estimator adapt by adjusting how much weight each instant contributes to the structured component. In particular, it allows the model to learn whether evidence nearer one end of the lookback window or the other carries greater informational value in the current window. This directional temporal decay is a departure from the original Gerber--IQ formalism, which discounts only into the past from the most recent observation. This flexibility is a modeling advantage, but it raises technical considerations for preserving PSD (especially under Hadamard/entrywise masking in atomic-IQ2), maintaining interpretational symmetry between opposite time--directions, and ensuring numerical stability.

To illustrate this we write $t=0,\dots,\tau\!-\!1$ to index the window from oldest to most recent, and define
\begin{equation}
a_t \;=\; \exp\!\Big(-\,\gamma\,(\tau-1-(t+\epsilon))_+\Big),
\qquad
(x)_+ := \max\{x,0\},
\label{eq:temporal-exponential}
\end{equation}
with delay $\epsilon\ge 0$. Interpreting $\gamma$ as the temporal-scaling parameter, $\gamma>0$ downweights older evidence, while $\gamma<0$ upweights it. For whole-atom scaling (atomic-IQ1), PSD is unaffected because $a_t>0$ simply rescales PSD atoms. However, more generally, two issues emerge: (i) we have interpretational asymmetry because transformations induced by $+\gamma$ and $-\gamma$ are not mirror images; and (ii) in off-diagonal scaling (atomic-IQ2), taking $v_t=a_t$ can yield $v_t>1$ when $\gamma<0$, violating the Schur bound $|v_t|\le 1$ thereby breaking the PSD guarantee under Hadamard masking.

To resolve both points, we separate magnitude and direction via a signed profile. Let the sign of $\gamma$ direct which edge of the window defines \emph{distance}, and let $|\gamma|$ control the decay rate in that direction. This gives,
\begin{equation}
\operatorname{age}_\gamma(t)
\;=\;
\begin{cases}
\tau-1-t, & \gamma \ge 0,\\[3pt]
t, & \gamma < 0,
\end{cases}
\qquad
a_t \;=\; \exp\!\Big(-\,|\gamma|\,(\operatorname{age}_\gamma(t)-\epsilon)_+\Big)\in(0,1].
\label{eq:signed-profile}
\end{equation}
With \eqref{eq:signed-profile}, atomic-IQ1 continues to scale entire atoms by $a_t$, preserving PSD. Atomic-IQ2 sets $v_t=a_t$, which automatically respects $|v_t|\le 1$. Additionally, positive/negative $\gamma$ now enjoy interpretational symmetry. The informative edge (recent vs.\ oldest) is now determined by $\operatorname{sign}(\gamma)$ while $|\gamma|$ sets the decay rate. The delay parameter $\epsilon$ creates a flat shelf at the chosen edge before decay begins.

For gradient-based estimation, one may smooth the kink at $\gamma=0$ without changing semantics, e.g.
\begin{equation}
\begin{aligned}
w(\gamma) &= \tfrac{1}{2}\bigl(1+\tanh(\kappa\gamma)\bigr) \\[6pt]
a_t &= \exp\!\Big(-\,\widetilde{|\gamma|}\,(\operatorname{age}_\gamma(t)-\epsilon)_+\Big) \\[6pt]
\operatorname{age}_\gamma(t) &= w(\gamma)(\tau-1-t) + \bigl(1-w(\gamma)\bigr)t \\[6pt]
\widetilde{|\gamma|} &= \sqrt{\gamma^2+\varepsilon_\gamma^2}
\end{aligned}
\label{eq:signed-smooth}
\end{equation}

with small $\varepsilon_\gamma>0$ and moderate $\kappa>0$, providing $C^\infty$ continuity at $\gamma=0$ and avoiding numerical artefacts. Modern gradient-based methods typically handle the unsmoothed kink without difficulty; the smoothed form is included out of caution rather than necessity.

In practice, atomic-IQ1 and atomic-IQ2 behave differently but coherently under this scheme. In the former, each event contributes $a_t A^{\pi_e}$, so diagonals (qualifying event) and off-diagonals (associated co-movement) are modulated together; uniform rescaling of $\{a_t\}$ cancels after normalization, only the shape of the temporal profile matters. In the latter, diagonals accumulate counts while off-diagonals are tempered by $a_t$. Increasing $|\gamma|$ downweights co-movement evidence, thereby moderating pairwise alignment relative to event occurrence, with the effect becoming stronger the further one moves from the chosen edge.

Two simple diagnostics make runs comparable across datasets and windows. The first is the effective mass,
\begin{equation}
\tau_{\mathrm{eff}} \;=\; \sum_{t=0}^{\tau-1} a_t,
\label{eq:meff}
\end{equation}
which measures the total weight assigned across the window. In atomic-IQ1 this captures how much of the window contributes to the structured component, while in atomic-IQ2 it indicates the relative strength with which co-movement is being tempered against raw activity. 

The second is the weighted mean age,
\begin{equation}
\overline{\mathrm{age}} \;=\; 
\frac{\sum_{t=0}^{\tau-1} a_t\,\operatorname{age}_\gamma(t)}
{\sum_{t=0}^{\tau-1} a_t},
\label{eq:age-mean}
\end{equation}
which locates the center of mass of the profile. In atomic-IQ1 it summarizes which portion of the window most influences $C^{(\alpha)}$, and in atomic-IQ2 it clarifies where correlations are most heavily adjusted relative to event occurrence. Reporting $(\tau_{\mathrm{eff}}, \overline{\mathrm{age}})$ alongside $(\gamma,\epsilon)$ makes runs interpretable and comparable across datasets and window lengths.

\section{Spectral Regimes}
\label{sec:SpectralRegimes}
The canonical squeezing identity

\begin{equation*}
S \;\equiv\; \sum_{\alpha \in \mathcal{K}} \eta_\alpha \, C^{(\alpha)}
\;+\;
\left( 1 - \sum_{\alpha \in \mathcal{K}} \eta_\alpha \right) I
\end{equation*}
expresses the estimator as a balance between the neutral benchmark \(I\) and the contributions of the channel classes. It is convenient to rewrite this in a compact form. Let
\[
\xi := \sum_{\alpha \in \mathcal{K}}\eta_\alpha, 
\qquad
P := \frac{1}{\xi}\sum_{\alpha \in \mathcal{K}}\eta_\alpha\,C^{(\alpha)} \quad (\xi>0).
\]
Then
\begin{equation}
\label{eq:impl-xi}
S \;=\; (1-\xi)I \;+\; \xi P .
\end{equation}
This representation shows that the estimator lies on the line segment between the neutral benchmark $I$ and the squeezing blend $P$ when $\xi \leq 1$, and continues beyond $P$ when $\xi > 1$. It is important to emphasize that neither $I$ nor $P$ are exogenous features; both are endogenous constructs intrinsic to the IQ framework. In particular, $P$ is not an external target but an internally generated average of channel-class matrices defined within the model.

\medskip

When the collective squeeze satisfies \(\xi \ll 1\), the estimator
\[
S \;=\; (1-\xi)I + \xi P \;\approx\; I + \xi (P - I)
\]
can be viewed as a perturbation of the identity benchmark by a small, structured deviation \(P - I\). In this regime, the IQ framework naturally enters a perturbative form, reminiscent of classical perturbation theory, where the response of a system is analyzed as a first-order correction to an unperturbed state. Here, \(P - I\) plays the role of a data-driven perturbation operator, and \(\xi\) controls its magnitude. The analogy highlights that the squeezing operation constitutes not merely an interpolation between two matrices, but a controlled perturbative deformation of \(I\) governed by interpretable, learned parameters.

\medskip

This observation opens a potentially fruitful line of interpretation: the perturbative squeezing regime. In this view, \(S\) may be regarded as the first term of a data-driven perturbation expansion, where higher-order corrections could, in principle, capture increasingly refined spectral interactions among the endogenous channel classes. Unlike conventional perturbation theory, the perturbation here is not exogenous but arises endogenously from the data itself, through the structured learning of the IQ parameters. Consequently, the matrix \(P\) can be interpreted as a tunable, positive semidefinite perturbation that embodies information-driven adjustments to the neutral benchmark. 

\medskip

This framing suggests that the Atomic-IQ architecture could be extended into a broader theory of data-driven perturbation of PSD operators. In portfolio optimization this perspective connects naturally with ideas of robustness and controlled deviation, where \(I\) represents neutrality or independence, and \(P\) encodes structured dependencies inferred from data. Although not developed further here, this conceptual link to perturbation theory may offer a promising analytical bridge between interpretable learning and spectral sensitivity analysis, and thus provides an avenue for future theoretical exploration.

Having introduced the conceptual interpretation of \(S\) as a perturbative blend of \(I\) and \(P\), we now turn to its spectral implications, which can be derived directly from the compact form in \eqref{eq:impl-xi}

The eigenvalues of \(S\) follow directly from \eqref{eq:impl-xi}. If \(\lambda_i(P)\) denotes the spectrum of \(P\), then
\[
\lambda_i(S) \;=\; (1-\xi) + \xi\,\lambda_i(P).
\]
It follows that
\[
S\succeq 0 \;\Longleftrightarrow\; (1-\xi)+\xi\,\lambda_{\min}(P)\geq 0.
\]
Two regimes emerge. When \(\xi\leq 1\), this inequality always holds provided the channel‑class matrices are positive semi‑definite, which they are in atomic-IQ, so positive semi‑definiteness is guaranteed. When \(\xi>1\), positive semi‑definiteness is no longer automatic and holds only if the smallest eigenvalue of \(P\) exceeds the threshold \(1-1/\xi\).

In general, the collective squeeze \(\xi\) is a free parameter, and choosing \(\xi<1\) is always an available option that guarantees positive semi‑definiteness regardless of the detailed structure of \(P\). In the original IQ paper \citep{gerberiqssrn}, however, we collapsed three class weights into a single parameter by making a specific allocation judgment: the tail class received full weight, the body class weight \(\eta\), and the wing class weight \(\eta^2\). This yields what we refer to as basic–IQ. A consequence of this parameterization is that the collective squeeze becomes
\[
\xi(\eta) = 1+\eta+\eta^2,
\]
which exceeds unity for any \(\eta>0\). Thus basic–IQ operates in the regime where positive semi‑definiteness is conditional. In earlier work this was handled by applying Higham’s (\cite{higham2002computing}) projection whenever the estimator was indefinite; with the spectral condition in hand we can now state explicitly when the basic–IQ estimator will be positive semi‑definite.

In the basic–IQ case the structured blend is
\[
P(\eta) = \frac{C^{(T)} + \eta\, C^{(B)} + \eta^{2}\, C^{(W)}}{1+\eta+\eta^{2}},
\]
so the condition becomes
\[
\lambda_{\min}\!\big(P(\eta)\big) \;\geq\; \frac{\eta+\eta^{2}}{1+\eta+\eta^{2}}.
\]
This criterion is necessary and sufficient. As $\eta$ approaches one, the weighting becomes uniform across classes and the template boundaries (which vanish at $\eta=1$) cease to affect the construction; the squeezed blend must move increasingly close to the identity in spectral terms to return a PSD matrix. More generally, as \(\eta\) increases from zero the requirement tightens smoothly.

In practice it can be useful to replace this exact test with bounds that depend only on the class eigenvalue minima. Let \(m_B=\lambda_{\min}(C^{(B)})\), \(m_W=\lambda_{\min}(C^{(W)})\), and \(m_T=\lambda_{\min}(C^{(T)})\). A simple sufficient condition is
\[
\eta^{2}(1-m_T)\;+\;\eta(1-m_W)\;\le\; m_B,
\]
while a necessary condition is
\[
\eta^{2}+\eta \;\le\; \frac{m_{\max}}{1-m_{\max}}, 
\qquad m_{\max}=\max\{m_B,m_W,m_T\}.
\]
These inequalities describe feasible ranges of \(\eta\) given information about the extremal eigenvalues of the class matrices. Proofs (with some routine steps omitted for brevity) for these spectral mapping and feasibility results, together with formulas for choosing the eigenfloor parameter to target a minimum eigenvalue or condition number, are given in the appendix.

\section{Eigenfloor and Conditioning}
\label{sec:eigenfloor}

The spectral analysis in the previous section distinguishes a guaranteed regime, where the collective squeeze \(\xi \le 1\), from a conditional regime, where \(\xi > 1\). In both settings it may be desirable to exercise explicit control over numerical conditioning. To this end we introduce an identity reserve, termed the \emph{eigenfloor}, which contracts the spectrum toward unity while remaining entirely within the squeezing framework.

\subsection{The Eigenfloor Mechanism}

The eigenfloor is defined for \(\phi \in [0,1]\) by
\begin{equation}
\label{eq:eigfloor-def}
\widehat S(\phi) \;=\; (1-\phi)\,S \;+\; \phi\,I.
\end{equation}
Substituting the canonical squeezing form \(S=\sum_\alpha \eta_\alpha C^{(\alpha)} + \bigl(1-\xi\bigr)I\) with \(\xi=\sum_\alpha \eta_\alpha\) yields
\[
\widehat S(\phi) \;=\; (1-\phi)\sum_\alpha \eta_\alpha C^{(\alpha)} \;+\; \bigl((1-\phi)(1-\xi)+\phi\bigr)I,
\]
which can be rearranged into the canonical representation with rescaled weights:
\begin{equation}
\label{eq:eigfloor-canonical}
\widehat S(\phi) \;=\; \sum_\alpha (1-\phi)\,\eta_\alpha\,C^{(\alpha)} \;+\; \Bigl(1-(1-\phi)\,\xi\Bigr)\,I.
\end{equation}
Equivalently, writing \(S=(1-\xi)I+\xi P = I + \xi(P-I)\),
\begin{equation}
\label{eq:eigfloor-xieff}
\widehat S(\phi) \;=\; I \;+\; \underbrace{(1-\phi)\,\xi}_{\xi_{\text{eff}}}\,(P-I),
\end{equation}
so the squeezing semantics are preserved: the class weights rescale to \(\eta'_\alpha=(1-\phi)\eta_\alpha\), and the effective collective squeeze reduces from \(\xi\) to \(\xi_{\text{eff}}=(1-\phi)\xi\).

Since \(I\) commutes with \(S\), the eigenvalues transform affinely:
\begin{equation}
\label{eq:eigfloor-map}
\lambda_i\bigl(\widehat S(\phi)\bigr) \;=\; (1-\phi)\,\lambda_i(S)+\phi.
\end{equation}
Each eigenvalue therefore moves toward \(1\). The spectral spread contracts: \(\lambda_{\min}\) is nondecreasing, \(\lambda_{\max}\) is nonincreasing, and for correlation-type matrices (trace \(=N\), where $N$ is the number of assets) the mean eigenvalue remains \(1\). Consequently, the floor simultaneously lowers the top of the spectrum and raises the bottom, improving numerical conditioning while leaving the total variance unchanged.

\subsection{Targeted Conditioning and PSD Repair}

Let \(\alpha=\lambda_{\max}(S)\) and \(\beta=\lambda_{\min}(S)\). Two closed-form rules allow one to choose \(t\) to achieve a desired spectral property.

\paragraph{Target minimum eigenvalue.}
For a specified \(\underline{\lambda}\in[0,1]\), equation~\eqref{eq:eigfloor-map} gives
\begin{equation}
\label{eq:t-for-lmin}
(1-\phi)\,\beta + \phi \;\ge\; \underline{\lambda}
\quad\Longleftrightarrow\quad
\phi \;\ge\; \frac{\underline{\lambda}-\beta}{\,1-\beta\,}\qquad(\beta<1).
\end{equation}
Clamping to the admissible range yields
\[
\phi_{\min}(\underline{\lambda}) \;=\; \Bigl[\;(\underline{\lambda}-\beta)/(1-\beta)\;\Bigr]_0^{\,1}.
\]
If \(\underline{\lambda}=0\) and \(\beta\ge 0\), no floor is required and one may set \(\phi=0\); if \(\beta<0\), then \(\phi_{\min}(0)=-\beta/(1-\beta)\in(0,1)\) is the smallest floor that restores positive semidefiniteness.

\paragraph{Target condition number.}
For a desired condition number \(K\ge 1\) and \(\beta>0\),
\begin{equation}
\label{eq:t-for-K}
\phi \;\ge\; \frac{\alpha-K\beta}{\,(\alpha-1)+K(1-\beta)\,}.
\end{equation}
If \(\beta\le 0\), first apply~\eqref{eq:t-for-lmin} to ensure \(\lambda_{\min}(\widehat S(\phi))>0\), then use~\eqref{eq:t-for-K} on the updated spectrum.  These expressions provide closed-form tuning rules for analytic PSD repair and explicit conditioning control without numerical optimization.

\subsection{Application in Conditional Regimes}
\label{sec:ConditionalRegimes}
In atomic–IQ each \(C^{(\alpha)}\) is positive semidefinite, implying \(P\succeq 0\) and \(\lambda_{\min}(P)\in[0,1]\).  When \(\xi>1\) and no information on \(\lambda_{\min}(P)\) is available, the bound \(\lambda_{\min}(S)\ge (1-\xi)+\xi\cdot 0 = 1-\xi\) implies that
\begin{equation}
\label{eq:t-worst}
\phi \;\ge\; 1-\frac{1}{\xi}
\end{equation}
guarantees \(\widehat S(\phi)\succeq 0\) irrespective of \(P\).  If a lower bound \(\lambda_{\min}(P)\ge \mu\in[0,1]\) is known, then \(\lambda_{\min}(S)\ge (1-\xi)+\xi\mu\) and the required floor can be reduced to
\begin{equation}
\label{eq:t-with-mu}
\phi \;\ge\; \max\Bigl\{\,0,\ 1-\frac{1}{\xi(1-\mu)}\,\Bigr\}.
\end{equation}
For the basic–IQ parameterization \(\xi(\eta)=1+\eta+\eta^2\), the conservative choice~\eqref{eq:t-worst} becomes \(\phi\ge 1-1/\bigl(1+\eta+\eta^2\bigr)\).

Although \eqref{eq:eigfloor-def} resembles shrinkage toward the identity, equations~\eqref{eq:eigfloor-canonical} and~\eqref{eq:eigfloor-xieff} show that the eigenfloor remains fully internal to the squeezing representation.  It constitutes a re-parameterization that rescales the class weights and increases the endogenous identity share.  Hence it performs two roles simultaneously: an analytic PSD-restoring mechanism when \(S\) is indefinite, and an explicit conditioning control when \(S\) is already PSD.

If \(S\) already meets the desired conditioning, set \(\phi=0\).  Small floors such as \(t\in[0.02,0.05]\) often stabilize optimization without materially altering class semantics.  When only \(\xi\) is known, the conservative bound~\eqref{eq:t-worst} ensures PSD; when partial spectral information on \(P\) is available, the refined bound~\eqref{eq:t-with-mu} is tighter and typically sufficient in practice.

\medskip
The analytic framework developed in Sections III–V establishes atomic–IQ as a constructive and spectrally controlled covariance estimator. By combining atomic building blocks, the canonical squeezing identity, and the eigenfloor mechanism, the estimator achieves both interpretability and explicit eigenvalue regulation. Having derived these properties in closed form, it is natural to ask how atomic–IQ behaves in practice when used within an economically meaningful optimization task. The next section addresses this question empirically, situating the estimator alongside established covariance models in a realistic portfolio-optimization setting.

\section{Empirical Study: Tangency Portfolio on a Multi-Asset Universe}
\label{sec:method}

This section demonstrates the atomic–IQ framework in a practical portfolio optimization setting. The aim is not exhaustive benchmarking, but to illustrate how the analytic properties established earlier, positive semi-definiteness by design, closed-form eigenvalue control, and interpretability, translate into realized portfolio behavior. Using a diversified multi-asset universe and a tangency (maximum-Sharpe) objective, atomic–IQ is evaluated alongside representative estimators from shrinkage, random-matrix, and concordance-based families. The empirical design emphasizes transparency and replicability: all estimators share the same data, constraints, and turnover assumptions, isolating the contribution of covariance estimation itself.

\subsection{Empirical Design}
\noindent\textbf{Purpose and scope.}
\label{sec:empericaldesign}
The preceding sections developed atomic–IQ (AIQ) as a constructive, PSD-guaranteed estimator offering analytic control over eigenvalues through the canonical squeezing identity and the eigenfloor mechanism. This section positions AIQ relative to established estimators within a single, economically meaningful objective, the long-only tangency (maximum-Sharpe) portfolio. The empirical study is intentionally compact yet complete, providing a clear comparison of estimator behavior in a realistic multi-asset environment. It details the data universe, backtesting protocol, competing estimators, and evaluation metrics, establishing a direct link between analytic properties and portfolio-level outcomes.

\vspace{0.35em}
\noindent\textbf{Asset universe and data.}
We reuse the 10‑asset universe from the IQ study (\cite{gerberiqssrn}): five equity indices (U.S.\ large‑cap, U.S.\ small‑cap, developed ex‑U.S., emerging markets, and U.S.\ growth), two bond indices (aggregate and high‑yield), listed real estate, gold, and a broad commodities index. We work with monthly total returns from January 1988 to December 2024 and rebalance monthly.

Many covariance estimators in the literature are motivated by high dimensional asymptotics, and their optimality results are typically derived for regimes in which the cross sectional dimension is large relative to the sample length. Relatedly, the random matrix perspective we adopt is underpinned by Marchenko Pastur type limits, which are most informative when applied to larger universes. We nevertheless focus on a realistic $N=10$ setting here, both for brevity and because ten asset universes remain common in strategic allocation and benchmark design. Readers interested in broader empirical evidence across alternative portfolio sizes and universes are referred to \cite{gerberiqssrn}; all results reported there for the Gerber IQ estimator can be replicated with Atomic IQ.

\vspace{0.35em}
\noindent\textbf{Backtest protocol.}
At each monthly decision date $d$ we estimate expected returns $\hat{\boldsymbol\mu}_d$ and the covariance matrix $\hat{\boldsymbol\Sigma}_d$ from a rolling lookback window of $\tau{=}20$ months, with $\hat{\boldsymbol\mu}_d$ computed as the sample mean over the same window. We then solve the long-only tangency problem
\[
\max_{\mathbf w\ge 0,\ \mathbf 1^\top\mathbf w=1}\ 
\frac{\mathbf w^\top \hat{\boldsymbol\mu}_d}{\sqrt{\mathbf w^\top \hat{\boldsymbol\Sigma}_d\,\mathbf w}},
\]
apply the resulting weights for the subsequent month, and repeat. Transaction costs of 10\,bps are charged on dollar turnover at each rebalance. The in-sample period (1988--1999) is used for hyperparameter selection, and out-of-sample performance is evaluated over January~2000--December~2024. When required, we enforce positive semi-definiteness using the analytic eigenfloor repair described above, preserving trace.

\noindent\textbf{Hyperparameter selection.} The atomic--IQ framework introduces a small number of design parameters that control how co-movement evidence is aggregated, including the alignment center \(r_0\), the volatility scaling scheme \(\sigma_{r_k}\) used for standardization (which sets the implied units for the noise gate \(c\) and the channel boundary \(\delta\)), the noise gate \(c\), the channel boundary \(\delta\), the channel squeezing weights $\boldsymbol{\eta}$, the temporal decay parameter \(\gamma\), the temporal decay delay parameter \(\epsilon\), the lookback duration parameter \(\tau\), and, when used, the eigenfloor level \(\phi\). Rather than tune these by hand, we treat them as hyperparameters and select them by data-driven search. To do so we employ Optuna, an open-source hyperparameter optimization library that implements adaptive sampling and early stopping strategies \citep{akiba2019optuna}, which allows us to explore the atomic--IQ parameter space efficiently without relying on a coarse grid search.

In the empirical study Optuna explores the ranges in Table~\ref{tab:aiq-params}, with the noise gate fixed at \(c = 0.5\) and the lookback duration fixed at \(\tau = 20\) months, which implies \(q = T/N = 2\) in the \(N=10\) asset setting. The parameters \(r_0\), \(\delta\), \(\eta\), \(\gamma\), and \(\epsilon\) are treated as free variables and optimised over the corresponding ranges in Table~\ref{tab:aiq-params}, with each candidate configuration evaluated by running the full tangency backtest on the in-sample period 1988--1999. Positive semi-definiteness is enforced using an eigenfloor \(\phi\), but \(\phi\) is not treated as a single tuned scalar; instead we adopt the fixed repair policy of Section \ref{sec:eigenfloor}.\ref{sec:ConditionalRegimes} and compute the realised floor \(\phi_d\) at each monthly covariance update. Specifically, with \(\xi\) denoting the collective squeeze and \(P\) the normalized structural component, we compute \(\mu = \lambda_{\min}(P)\) and set \(\phi_d = 0\) whenever the feasibility condition holds, otherwise we apply the tight analytic floor \(\phi_d = \max\{0,\, 1 - [\xi(1-\mu)]^{-1}\}\) before forming the repaired estimator. For a given Optuna trial we then construct the corresponding AIQ1 or AIQ2 covariance estimator, solve the long-only tangency problem with 10~bp transaction costs at each monthly rebalance, and record the resulting after-cost Sharpe ratio. Optuna uses these scalar evaluations to concentrate the search in promising regions of the parameter space, delivering well-behaved parameter sets with a modest computational budget. The best-performing configuration for each member of the atomic--IQ family (AIQ1 and AIQ2) is then held fixed and carried forward into the out-of-sample evaluation from 2000 to 2024. To support reproducibility, we provide a reference implementation of the full pipeline in a public repository.\footnote{\url{https://doi.org/10.5281/zenodo.18069453}}

\vspace{0.35em}
\noindent\textbf{Estimators compared.}

We benchmark atomic–IQ against shrinkage, random‑matrix theory, and concordance‑based constructions, plus a historical sample covariance (HC) baseline. Table~\ref{tab:1} lists the families referred to in the results.\\

\begingroup
\footnotesize
\setlength{\LTpre}{0pt}
\setlength{\LTpost}{0pt}
\begin{longtable}{>{\raggedright\arraybackslash}p{3.5cm} >{\raggedright\arraybackslash}p{10.0cm}}
\toprule
\textbf{Technique} & \textbf{Description} \\
\midrule
\endfirsthead

\toprule
\textbf{Technique} & \textbf{Description} \\
\midrule
\endhead

\multicolumn{2}{l}{\textbf{Shrinkage}} \\
Linear (LS1) &
Shrinkage toward a one-parameter matrix: all variances are equal, and all covariances are zero \citep{ledoit2004well}. \\
Linear (LS2) &
Shrinkage toward a two-parameter matrix: all variances are equal, and all covariances are equal \citep{ledoit1995essays}. \\
Linear (LS3) &
Shrinkage toward a constant-correlation matrix: the target matrix preserves the diagonal of the sample covariance matrix and sets all correlations equal \citep{ledoit2004honey}. \\
Linear (LS4) &
Shrinkage toward a diagonal matrix: the target matrix preserves the sample variances and sets covariances to zero \citep{ledoit1995essays}. \\
Linear (LS5) &
Shrinkage toward a one-factor market model where the factor is the cross-sectional average of the variables; variances are preserved \citep{ledoit2003improved}. \\
Non-linear (NLS1) &
Geometric-inverse shrinkage under symmetrized Kullback-Leibler loss; averages linear-inverse and quadratic-inverse shrinkage \citep{ledoit2022quadratic}. \\
Non-linear (NLS2) &
Linear-inverse shrinkage derived under Stein's loss \citep{ledoit2022quadratic}. \\
Non-linear (NLS3) &
Quadratic-inverse shrinkage derived under Frobenius loss, inverse Stein's loss, and minimum variance loss \citep{ledoit2022quadratic}. \\

\midrule
\multicolumn{2}{l}{\textbf{Random Matrix Theory}} \\
Constant Residual Eigenvalue (CRE) &
Applies the Marchenko-Pastur theorem to identify noise-associated eigenvalues, replacing them with their mean to preserve the trace \citep{de2020machine}. \\
Shrinkage of Residual Eigenvalues (SRE) &
Similar to CRE, but shrinks noise eigenvalues toward a diagonalized form while preserving the trace \citep{de2020machine}. \\

\midrule
\multicolumn{2}{l}{\textbf{Co-Movement}} \\
Gerber Statistic (GS) &
Inspired by Kendall’s Tau, uses a noise exclusion zone, scaled axes and non-translated origin (\(r_i=0,r_j =0\)) \citep{gerber2022gerber}. \\
Atomic-IQ (AIQ) &
Inspired by the Gerber IQ statistic, a PSD squeezing co-movement statistic with analytic control of eigenvalues and condition number\\
\bottomrule
\caption{Collection of candidate techniques for covariance matrix estimation against which atomic squeezing is tested. Testing is also performed against the untreated sample covariance matrix.}
\label{tab:1}
\end{longtable}
\endgroup

\vspace{0.35em}
\noindent\textbf{Atomic--IQ parameterization and PSD control.}
Atomic--IQ uses spatial and temporal controls that (i) align marginal distributions, (ii) identify body/tail/wing channels via thresholds, and (iii) assign information-weighted squeezing. Table \ref{tab:aiq-params} summarizes the parameters along with the typical ranges used in the in-sample optimization.

PSD is guaranteed either (i) by design when the collective squeeze respects the canonical identity (no eigenvalues cross zero), or (ii) by applying an analytic eigenfloor $\lambda_k\leftarrow\max\{\lambda_k,\phi\}$ to the spectrum of the squeezed matrix, for a small $\phi\!>\!0$ chosen within the squeezing framework; both routes are optimization-friendly and maintain trace consistency.

\medskip
\noindent\textbf{Atomic--IQ squeezing weights.}
In \emph{free} mode, which is what we use in this study, Atomic--IQ parameterizes the three channel weights via independent logits $\mathbf z=(z_B,z_W,z_T)$ mapped through the sigmoid, $\eta_k=\sigma(z_k)$ for $k\in\{B,W,T\}$, with $\sigma(x)=1/(1+e^{-x})$. The zero-correlation benchmark weight is then determined residually as
\[
1-\xi \;=\; 1 - (\eta_B+\eta_W+\eta_T).
\]
In our calibrated solutions (see Table \ref{tab:aiq-final-params}) $\xi>1$, placing the estimator in the conditional regime; the framework and analytic PSD controls are designed to accommodate this while maintaining PSD status of the overall covariance estimator $\hat{S}$.

\noindent\textbf{Optuna budget and implications of the calibrated parameters.}
The Atomic–IQ Optuna hyperparameter search is run for 5{,}000 trials (candidate parameter configurations). This is a deliberately modest search budget, chosen as a compromise between computational cost and accessibility, so that the calibration protocol can be replicated on limited hardware. At the same time, 5{,}000 trials remains a relatively limited exploration for multi-parameter estimators, particularly Atomic--IQ, where interactions across spatial and temporal controls can create a high-dimensional search landscape.

Inspecting the calibrated solutions, both Atomic--IQ1 and Atomic--IQ2 select $\epsilon\approx 19$ (to the nearest integer), which corresponds to the upper bound $\tau-1=19$ under a $\tau=20$ month lookback. Since $\epsilon$ governs temporal discounting, this indicates that, in this study, the optimal degree of temporal decay is effectively zero, with past co-movement contributions retained at full temporal weight. Consequently, the reported values of $\gamma$ should be interpreted as artifacts of the constrained finite search rather than as evidence of a meaningful optimal temporal-discounting mechanism in the present setting. This temporal conclusion is independent of the spatial allocation, which is controlled by the channel weights.

Finally, the calibrated channel weights imply distinct spatial emphasis across the two variants. For Atomic--IQ1 the ordering is $\eta_T > \eta_B > \eta_W$, indicating strongest emphasis on tail co-movements, followed by body, with the wing channel receiving the smallest of the three channel weights. For Atomic--IQ2 the ordering is $\eta_B \approx \eta_T \gg \eta_W$, indicating an almost symmetric allocation between body and tail with a negligible wing component.

\medskip
\noindent\textbf{Parameterization of tunable benchmarks.}
Among the alternative estimators, only the Gerber Statistic (GS) and the random-matrix shrinkage estimator (SRE) are directly tunable in our study. For GS, the tunable hyperparameter is the threshold $\theta$ that defines the co-movement event regions. We report results for $\theta\in\{0.5,\,0.7,\,0.9\}$ to match the settings used in the original Gerber study, and we additionally consider an optimized setting $\theta^{\ast}=0.439$ obtained via the same Optuna in-sample optimization protocol used for Atomic--IQ. We denote this optimized variant as GS$^{\ast}$.

For SRE, the tunable hyperparameter is the shrinkage-mix weight $\alpha$, which controls the weight assigned to the sample covariance matrix in the shrinkage combination. We report SRE with $\alpha=0.1$, which is commonly used in related work, and an optimized variant SRE$^{\ast}$ where $\alpha^{\ast}$ is selected using the same Optuna protocol.

Strictly, both SRE and CRE also require a kernel density estimation bandwidth; throughout we fix this bandwidth at $h=0.01$ for both methods, following common practice, and we do not tune $h$. All Optuna-based hyperparameter selections are conducted using the in-sample calibration period (1988--1999) only.

\FloatBarrier
\begingroup
\setlength{\tabcolsep}{5pt} 
\begin{table}[!htbp]
\centering
\footnotesize
\renewcommand{\arraystretch}{1.05}
\begin{tabular}{@{} l p{8.8cm} p{3.7cm} @{}}
\toprule
\textbf{Parameter} & \textbf{Interpretation} & \textbf{Typical Values} \\
\midrule
\multicolumn{3}{l}{\textbf{Spatial Parameters}} \\
\texttt{$r_0$} (center) & Establishes a co-movement center for the delineation of concordant and discordant co-movement, a critical component of statistical distributional alignment. & $r_0 \in \{ \bar{r}, \tilde{r}, 0 \}$\textsuperscript{\dag}
 \\
$c$ (exclusion) & Reduces noise by filtering out observations with low co-movement informational value. Given in units of the scaling parameter $\sigma_{r_k}$. &
\begin{tabular}[t]{@{}l@{}}
fixed at 0.5 in this study
\end{tabular} \\
$\delta$ (boundary) & Establishes squeezing channel boundaries that separate moderate from extreme returns in scaled return space, enabling empirical tail detection. Given in units of the scaling parameter $\sigma_{r_k}$. & 1 - 3 \\
$\sigma_{r_k}$(scaling) & Implements volatility scaling used to map returns into scaled return space when applying \(c\) and \(\delta\). The choice affects interpretation of squeezing channels. &
\begin{tabular}[t]{@{}l@{}}
\( f(\hat{\sigma}_{r_i}, \hat{\sigma}_{r_j}) \in \) \\
\(\Big\{(\hat{\sigma}_{r_i}, \hat{\sigma}_{r_j}),\;
\max(\hat{\sigma}_{r_i}, \hat{\sigma}_{r_j}),\;\Big.\) \\
\(\Big.\min(\hat{\sigma}_{r_i}, \hat{\sigma}_{r_j}),\;
\operatorname{mean}(\hat{\sigma}_{r_i}, \hat{\sigma}_{r_j})\Big\}\)
\end{tabular} \\
$\eta$ (squeeze) & Controls the squeezing intensity assigned to observations within defined channels. Noisier data are allocated lower values (closer to 0), indicating stronger squeezing, while more informative data receive values closer to 1, preserving their influence. & 0 – 1 \\
\midrule
\multicolumn{3}{l}{\textbf{Temporal Parameters}} \\
$\tau$ (duration) & Lookback window length for co-movement estimation. Balances responsiveness with statistical and computational stability. & fixed at 20 months to give $q=\tau/N=2$ \\
$\varepsilon$ (delay) & Implements a flat-weighting window to allow data to retain full influence prior to the onset of temporal decay. & $
\varepsilon \in [0,\tau-1]$ \\
$\gamma$ (decay) & Controls temporal weighting; \( |\gamma| \) sets the discount strength and the sign of \(\gamma\) determines whether emphasis is placed toward the start or the end of the lookback window. The discount rate is parameterized via the half-life \(T_{1/2}=\ln(2)/|\gamma|\), with \(T_{1/2}=\infty\) corresponding to \(\gamma=0\). & $T_{1/2}\in[6.93,\infty)\;\equiv\;|\gamma|\in[0,0.1]$ \\
\bottomrule
\end{tabular}
\caption{Parameter guidance: grouped by function with interpretation and typical values. \textsuperscript{\dag}$\{\bar r, \tilde r, 0\}$ denotes, respectively, the sample mean return, the sample median return, and zero return.}
\label{tab:aiq-params}
\end{table}
\endgroup

\begin{table}[t]
\centering
\begin{tabular}{lcc}
\hline
Parameter & Atomic--IQ1 & Atomic--IQ2 \\
\hline
$r_0$ & 0 & 0 \\
$\sigma_{r_k}$ & $\max\!\left(\hat{\sigma}_{r_i},\,\hat{\sigma}_{r_j}\right)$ & $\max\!\left(\hat{\sigma}_{r_i},\,\hat{\sigma}_{r_j}\right)$ \\
$\delta$ & 1.50 & 1.52 \\
$\eta_B$ & 0.956 & 0.976 \\
$\eta_W$ & 0.870 & 0.0190 \\
$\eta_T$ & 0.979 & 0.976 \\
$\gamma^\dagger$ & 0.0541 & 0.0765 \\
$\epsilon$ & 19.0 & 18.9 \\

\hline
\end{tabular}
\caption{In-sample calibrated Atomic--IQ settings used in the out-of-sample backtests.\,\textdagger\ Note that \(\varepsilon\) (delayed decay) takes close to its maximum permissible value, rendering the reported values of \(\gamma\) artefacts; in effect there is no temporal discounting.
}
\label{tab:aiq-final-params}
\end{table}

\vspace{0.35em}
\noindent\textbf{Evaluation metrics.}
We report the annualized Sharpe ratio, the Sortino ratio, the Calmar ratio, the annualized return, the cumulative return, the annualized volatility ($\sigma$), the maximum drawdown, the 95\% monthly value-at-risk, and the average monthly turnover. The out-of-sample window for all tables is 2000--2024.

We also report statistical significance tests for out-of-sample performance differences, with particular emphasis on pairwise tests of out-of-sample Sharpe ratio differences between atomic--IQ and each competing estimator.

\bigskip
\subsection{Empirical Results}
\label{sec:empirical}
Table~4 reports out-of-sample performance for all covariance estimators under the common long-only tangency protocol described in Section~VI.A (monthly rebalancing, $\tau=20$ months for both $\hat{\boldsymbol\mu}$ and $\hat{\boldsymbol\Sigma}$, and 10\,bp transaction costs applied to dollar turnover). Results are ordered by decreasing after-cost Sharpe ratio.

Several patterns are immediate. First, the atomic--IQ variants sit at the top of the Sharpe distribution: AIQ1 delivers the highest out-of-sample Sharpe ratio (0.56), with AIQ2 close behind (0.54). This risk-adjusted performance is achieved with comparatively low volatility (7.70\% for AIQ1 and 7.65\% for AIQ2), and with moderate drawdowns (approximately $-28\%$) relative to the higher-volatility shrinkage alternatives. In contrast, the best-performing linear shrinkage variant by Sharpe, LS1 (0.54), attains materially higher annualized return, but it does so with substantially higher volatility (11.40\%), deeper drawdowns ($-36.53\%$), and a more adverse 95\% VaR (Table~4).

Second, methods that emphasize raw return, such as LS2, rank highly on annualized and cumulative return but are penalized by markedly higher risk. LS2 achieves the highest annualized and cumulative returns in the table, yet it exhibits the highest volatility and one of the worst drawdowns and VaR outcomes, which depresses its Sharpe ratio and makes its risk profile less attractive for a tangency objective.

Third, the concordance-based Gerber family is competitive but does not dominate: GS1 and the Optuna-tuned GS$^{\ast}$ deliver similar Sharpe ratios (both 0.51), with broadly comparable tail-risk statistics, while GS3 is slightly weaker (0.50) and GS2 is clearly the weakest among the Gerber variants (0.44). This ordering is consistent with the view that concordance-based construction is effective in this setting, but that atomic squeezing can improve risk-adjusted performance while also providing PSD guarantees and explicit spectral control.

Finally, the random-matrix cleaners (CRE and SRE) do not improve Sharpe in this particular $N=10$ universe. SRE and SRE$^{\ast}$ cluster around the historical covariance baseline (HC), and CRE is slightly lower. Overall, the table suggests that AIQ1 achieves its Sharpe premium primarily through a more favorable risk profile (lower volatility and milder tail risk) rather than through unusually high raw returns, and it does so without incurring unusually high turnover relative to competing approaches.

\subsection{Overall Rankings Across Estimators}
\label{sec:results}

To summarize performance beyond Sharpe alone, Table~5 reports per-metric ranks for every estimator across the full set of reported outcomes (Sharpe, Sortino, Calmar, annualized return, cumulative return, volatility, maximum drawdown, 95\% VaR, and turnover). The aggregate score is the sum of these ranks (lower is better), providing a compact measure of broad risk--return quality under the common backtest protocol.

The aggregate rankings reinforce the central message of Table~4. AIQ1 attains the best overall score (34), reflecting a combination of first-ranked Sharpe performance and consistently strong risk metrics, particularly volatility and tail-risk measures, without sacrificing turnover. AIQ2 ranks second overall (52), driven by similarly low volatility and strong Sharpe and turnover ranks, though with weaker return ranks than some shrinkage competitors. The next cluster comprises the Gerber variants GS1 and GS$^{\ast}$ (aggregate scores in the low-to-mid 60s), which benefit from strong drawdown and Calmar ranks, but do not match atomic--IQ on Sharpe and volatility simultaneously.

Several trade-offs become clearer in the rank table. LS1 ranks near the top on Sharpe and performs strongly on return, but its high volatility and adverse tail metrics push it down in aggregate rank. LS2 illustrates an even starker version of this pattern: it is first-ranked on annualized and cumulative return, yet it ranks last (or near-last) on volatility, drawdown, and VaR, yielding a poor aggregate score despite headline return strength. Conversely, the sample covariance (HC) and the RMT cleaners (SRE, SRE$^{\ast}$, CRE) occupy the middle-to-lower portion of the aggregate ordering, indicating that in this low-dimensional setting their improvements in stability do not translate into superior out-of-sample tangency performance once costs and tail risk are accounted for.

Taken together, Tables~4--5 indicate that atomic--IQ, and AIQ1 in particular, offers a favorable combination of high risk-adjusted performance and broad risk control, rather than a narrow improvement in a single metric. This is consistent with the intended role of the squeezing framework, namely to produce covariance estimates that are optimization-ready and spectrally controlled while still improving realized portfolio behavior.

\renewcommand{\arraystretch}{1.08}
\begin{table}[!t]
\centering
\footnotesize
\setlength{\tabcolsep}{3.9pt}
\begin{threeparttable}
\begin{tabular}{
  l
  S[table-format=1.2]      
  S[table-format=2.2]      
  S[table-format=3.2]      
  S[table-format=2.2]      
  S[table-format=-2.2]     
  S[table-format=-1.2]     
  S[table-format=1.2]      
  S[table-format=1.2]      
  S[table-format=1.2]      
}
\toprule
\multicolumn{1}{c}{Estimator} & \multicolumn{1}{c}{SR} & \multicolumn{1}{c}{AR} & \multicolumn{1}{c}{CR} & \multicolumn{1}{c}{$\sigma$} & \multicolumn{1}{c}{MDD} & \multicolumn{1}{c}{VaR95} & \multicolumn{1}{c}{So} & \multicolumn{1}{c}{Cal} & \multicolumn{1}{c}{Turn} \\
\midrule
AIQ1 & 0.56 & 5.91 & 267.93 & 7.70 & -28.06 & -3.16 & 0.68 & 0.21 & 2.81 \\
AIQ2 & 0.54 & 5.69 & 249.98 & 7.65 & -28.29 & -3.28 & 0.65 & 0.20 & 2.76 \\
LS1 & 0.54 & 7.71 & 412.45 & 11.40 & -36.53 & -4.71 & 0.65 & 0.21 & 3.32 \\
LS2 & 0.52 & 8.51 & 484.81 & 13.22 & -41.75 & -5.90 & 0.64 & 0.20 & 3.96 \\
NLS6 & 0.52 & 6.53 & 308.82 & 9.60 & -30.22 & -4.04 & 0.62 & 0.22 & 3.69 \\
NLS8 & 0.52 & 6.58 & 312.49 & 9.66 & -30.21 & -4.04 & 0.62 & 0.22 & 3.70 \\
GS1 & 0.51 & 5.91 & 262.04 & 8.52 & -27.22 & -3.35 & 0.60 & 0.22 & 3.47 \\
GS* & 0.51 & 5.92 & 262.77 & 8.56 & -27.22 & -3.35 & 0.60 & 0.22 & 3.47 \\
NLS7 & 0.51 & 6.48 & 304.65 & 9.55 & -30.25 & -4.03 & 0.62 & 0.21 & 3.69 \\
GS3 & 0.50 & 5.59 & 239.90 & 8.01 & -29.36 & -3.51 & 0.60 & 0.19 & 3.31 \\
HC & 0.49 & 5.94 & 261.90 & 8.89 & -30.50 & -3.39 & 0.59 & 0.19 & 3.45 \\
LS3 & 0.49 & 6.14 & 275.83 & 9.33 & -32.13 & -3.73 & 0.59 & 0.19 & 3.41 \\
SRE* & 0.49 & 5.94 & 261.86 & 8.89 & -30.50 & -3.39 & 0.59 & 0.19 & 3.45 \\
LS5 & 0.48 & 5.78 & 250.25 & 8.65 & -31.45 & -3.06 & 0.58 & 0.18 & 3.34 \\
SRE & 0.48 & 5.82 & 252.96 & 8.74 & -30.55 & -3.26 & 0.58 & 0.19 & 3.46 \\
CRE & 0.47 & 5.74 & 245.91 & 8.84 & -32.10 & -3.07 & 0.56 & 0.18 & 3.45 \\
LS4 & 0.47 & 5.53 & 232.19 & 8.50 & -30.01 & -3.32 & 0.56 & 0.18 & 3.23 \\
GS2 & 0.44 & 5.30 & 215.24 & 8.37 & -29.32 & -3.65 & 0.52 & 0.18 & 3.44 \\
\bottomrule
\end{tabular}
\caption{Out of sample performance across all estimators and metrics, ordered on decreasing Sharpe ratio; monthly rebalancing; 10\,bp costs.}
\label{tab:family_best_clean_rewrite}
\end{threeparttable}
\end{table}

\renewcommand{\arraystretch}{1.08}
\begin{table}[!t]
\centering
\footnotesize
\setlength{\tabcolsep}{3.9pt}
\begin{threeparttable}
\begin{tabular}{
  l
  S[table-format=2.0]  
  S[table-format=2.0]  
  S[table-format=2.0]  
  S[table-format=2.0]  
  S[table-format=2.0]  
  S[table-format=2.0]  
  S[table-format=2.0]  
  S[table-format=2.0]  
  S[table-format=2.0]  
  S[table-format=3.0]  
}
\toprule
\multicolumn{1}{c}{Estimator} &
\multicolumn{1}{c}{SR} &
\multicolumn{1}{c}{AR} &
\multicolumn{1}{c}{CR} &
\multicolumn{1}{c}{$\sigma$} &
\multicolumn{1}{c}{MDD} &
\multicolumn{1}{c}{VaR95} &
\multicolumn{1}{c}{So} &
\multicolumn{1}{c}{Cal} &
\multicolumn{1}{c}{Turn} &
\multicolumn{1}{c}{Agg} \\
\midrule
AIQ1     & 1  & 10 & 7  & 2  & 3  & 3  & 1  & 5  & 2  & 34  \\
AIQ2     & 2  & 15 & 14 & 1  & 4  & 5  & 2  & 8  & 1  & 52  \\
GS1      & 7  & 10 & 9  & 6  & 1  & 7  & 9  & 1  & 13 & 63  \\
GS*      & 7  & 9  & 8  & 7  & 1  & 10 & 8  & 1  & 14 & 65  \\
LS1      & 2  & 2  & 2  & 17 & 17 & 17 & 2  & 5  & 5  & 69  \\
NLS6     & 4  & 4  & 4  & 15 & 10 & 15 & 5  & 1  & 15 & 72  \\
NLS8     & 4  & 3  & 3  & 16 & 9  & 15 & 5  & 1  & 17 & 72  \\
NLS7     & 7  & 5  & 5  & 14 & 11 & 14 & 5  & 5  & 15 & 80  \\
GS3      & 10 & 16 & 16 & 3  & 6  & 11 & 9  & 10 & 4  & 84  \\
HC       & 11 & 7  & 10 & 11 & 12 & 9  & 11 & 10 & 9  & 88  \\
SRE*     & 11 & 7  & 11 & 10 & 12 & 9  & 11 & 10 & 9  & 89  \\
LS2      & 4  & 1  & 1  & 18 & 18 & 18 & 4  & 8  & 18 & 90  \\
LS3      & 11 & 6  & 6  & 13 & 16 & 13 & 11 & 10 & 7  & 93  \\
LS5      & 14 & 13 & 13 & 8  & 14 & 1  & 14 & 15 & 6  & 98  \\
SRE      & 14 & 12 & 12 & 9  & 15 & 4  & 14 & 10 & 12 & 100 \\
LS4      & 16 & 17 & 17 & 5  & 8  & 6  & 16 & 15 & 3  & 102 \\
CRE      & 16 & 14 & 15 & 12 & 17 & 2  & 16 & 15 & 9  & 112 \\
GS2      & 18 & 18 & 18 & 4  & 5  & 12 & 18 & 15 & 8  & 116 \\
\bottomrule
\end{tabular}
\caption{Aggregate ranking of out of sample performance across all estimators and metrics (lower is better.)}
\label{tab:overall_ranking_clean}
\scriptsize
\end{threeparttable}
\end{table}

\subsection{Sharpe Ratio Comparison and Statistical Inference}
Table~4 reports the out-of-sample performance of all competing covariance estimators under the common long-only tangency protocol, and Table~5 aggregates ranks across the full set of reported metrics. AIQ1 attains the highest after-cost Sharpe ratio (0.56) among all methods (Table~4), but the key feature is that this Sharpe is delivered with a distinctly favourable overall risk profile. Relative to the main conventional competitors that are closest on Sharpe, such as LS1 (0.54), AIQ1 operates at materially lower volatility (7.70\% versus 11.40\%), with milder drawdowns and tighter tail-risk, while also exhibiting lower turnover (Table~4). This broad stability is reflected in the aggregate ranking, where AIQ1 achieves the best overall score (34), followed by AIQ2 (52), with the remaining estimators spanning larger totals (Table~5). In this sense, AIQ1 does not appear to obtain a Sharpe premium by accepting hidden risk elsewhere in the distribution.

Because differences in Sharpe ratios across the top of the table are modest, we also ask whether any alternative estimator can be shown to deliver a Sharpe ratio that is statistically higher than AIQ1's. Treating AIQ1 as the benchmark, we test one-sided hypotheses of the form $H_0:\, SR(k)\ge SR(\mathrm{AIQ1})$ versus $H_1:\, SR(k)<SR(\mathrm{AIQ1})$ for each competitor $k$ (Appendix~C). In full-sample terms, using a moving-block bootstrap over monthly excess returns, the evidence supports rejection of $H_0$ for the weaker methods, with GS2 and LS4 in particular exhibiting Sharpe ratios that are significantly lower than AIQ1 at conventional levels. For the remaining estimators, the bootstrap confidence intervals for the full-sample Sharpe difference include zero, so there is no statistical evidence that any method improves on AIQ1's Sharpe once estimation error and time dependence are accounted for.

To assess stability of relative Sharpe performance through time, we compute 36-month rolling Sharpe ratios and form the difference series
\[
D^{(k)}_t = SR_t(\mathrm{AIQ1}) - SR_t(k),
\]
for each competitor $k$, evaluated both on the monthly grid and on a thinned grid with a step of three months to reduce overlap across windows (Appendix~C). On the monthly grid, the average rolling Sharpe difference is positive against every competing estimator, with typical mean differentials in the range 0.04 to 0.16 in annualised units. Accounting for strong serial dependence induced by overlapping windows, Newey--West HAC inference and a moving-block bootstrap for the mean difference again identify GS2 and LS4 as robustly inferior to AIQ1 across both grids. For several of the stronger competitors, including LS1 and the nonlinear shrinkage estimators, the HAC p-values on the monthly grid are small, suggesting that AIQ1's rolling Sharpe advantage may be economically meaningful even when statistical separation is not uniformly sharp under bootstrap uncertainty. Overall, the inference results support the practical conclusion suggested by Tables~4--5: several estimators are Sharpe-comparable to AIQ1, none can be shown to outperform it on Sharpe, and the weakest alternatives are clearly dominated.

\section{Conclusion}

This paper develops squeezed covariance estimation, a constructive framework that advances Informational Quality by resolving two core obstacles that routinely limit the practical use of concordance-based covariance estimators. First, it provides PSD guarantees by construction. Channel-class matrices are assembled from positive semi-definite atoms and normalized at the class level, so the estimator remains PSD whenever the collective squeeze lies in the feasible region. For the conditional regime, we derive an exact feasibility condition and illustrate it under the basic--IQ parameterization, making the PSD boundary explicit rather than relying on ex-post projection. Second, the framework delivers analytic eigenvalue control. We introduce an eigenfloor mechanism that enforces a positive spectral margin and supplies closed-form rules for targeting either a minimum-eigenvalue floor or a desired condition number. Importantly, the same mechanism functions as an analytic PSD repair in the conditional regime while remaining expressible within the canonical squeezing identity. The resulting estimator is interpretable, spectrally disciplined, and immediately optimization-ready, offering a principled alternative to ad hoc PSD fixes and placing concordance-based methods on equal footing with sample-anchored regularization.

Empirically, under a common long-only tangency backtest with transaction costs, atomic--IQ delivers the strongest risk-adjusted performance when compared against the full set of estimators in the study. The main economic message is not merely a higher Sharpe ratio, but a consistently improved risk profile: atomic--IQ achieves its Sharpe advantage with lower realized volatility and tighter tail-risk characteristics, rather than by accepting hidden fragility elsewhere in the return distribution. This is precisely the type of improvement that matters for implementable portfolio choice, where stability and robustness are as important as headline performance.

Finally, the squeezing construction is dimension-agnostic and naturally accommodates changes in the portfolio dimension $N$ and the sampling ratio $Q=T/N$. This makes the framework relevant across both data-rich and data-poor regimes, including settings where classical sample-based covariance estimation becomes unstable and where PSD feasibility and spectral control are most consequential.

\bibliographystyle{apalike}
\bibliography{refs} 

@article{akiba2019optuna,
  title  = {{Optuna}: A Next-Generation Hyperparameter Optimization Framework},
  author = {Akiba, Takuya and Sano, Shotaro and Yanase, Toshihiko and Ohta, Takeru and Koyama, Masanori},
  year   = {2019},
  note   = {arXiv:1907.10902},
  url    = {https://arxiv.org/abs/1907.10902}
}

@article{bun2017cleaning,
  title     = {Cleaning Large Correlation Matrices: Tools from Random Matrix Theory},
  author    = {Bun, Jo{\"e}l and Bouchaud, Jean-Philippe and Potters, Marc},
  journal   = {Physics Reports},
  volume    = {666},
  pages     = {1--109},
  year      = {2017},
  publisher = {Elsevier}
}

@article{chamberlain1983factor,
  title     = {Arbitrage, Factor Structure, and Mean-Variance Analysis on Large Asset Markets},
  author    = {Chamberlain, Gary and Rothschild, Michael},
  journal   = {Econometrica},
  volume    = {51},
  pages     = {1281--1304},
  year      = {1983},
  publisher = {Wiley}
}

@article{fan2013factor,
  title     = {Large Covariance Estimation by Thresholding Principal Orthogonal Complements},
  author    = {Fan, Jianqing and Liao, Yuan and Mincheva, Martina},
  journal   = {Journal of the Royal Statistical Society: Series B (Statistical Methodology)},
  volume    = {75},
  pages     = {603--680},
  year      = {2013},
  publisher = {Wiley}
}

@article{gerber2022gerber,
  title   = {The Gerber Statistic: A Robust Co-Movement Measure for Portfolio Optimization},
  author  = {Gerber, Sander and Markowitz, Harry M. and Ernst, Philip A. and Miao, Yinsen and Javid, Babak and Sargen, Paul},
  journal = {The Journal of Portfolio Management},
  volume  = {48},
  year    = {2022}
}

@article{gerberiqssrn,
  title   = {Squeezing Financial Noise: A Novel Approach to Covariance Matrix Estimation},
  author  = {Gerber, Sander and Smyth, William S. and Markowitz, Harry M. and Ernst, Philip A. and Miao, Yinsen and Sargen, Paul},
  year    = {2025},
  journal = {SSRN Electronic Journal},
  doi     = {10.2139/ssrn.4986939},
  url     = {https://ssrn.com/abstract=4986939},
  note    = {SSRN Working Paper No. 4986939}
}

@article{minns2025ssrn,
  title   = {Markowitz-Informed Neural Networks ({MINNs}): An Interpretable Deep Learning Approach to Portfolio Optimization},
  author  = {Smyth, William S. and Ernst, Philip A. and Miao, Yinsen},
  year    = {2025},
  journal = {SSRN Electronic Journal},
  doi     = {10.2139/ssrn.5336779},
  url     = {https://ssrn.com/abstract=5336779},
  note    = {SSRN Working Paper No. 5336779}
}

@article{higham2002computing,
  title     = {Computing the Nearest Correlation Matrix, a Problem from Finance},
  author    = {Higham, Nicholas J.},
  journal   = {IMA Journal of Numerical Analysis},
  volume    = {22},
  pages     = {329--343},
  year      = {2002},
  publisher = {Oxford University Press}
}

@article{laloux1999,
  title     = {Noise Dressing of Financial Correlation Matrices},
  author    = {Laloux, Laurent and Cizeau, Pierre and Bouchaud, Jean-Philippe and Potters, Marc},
  journal   = {Physical Review Letters},
  volume    = {83},
  pages     = {1467--1470},
  year      = {1999},
  publisher = {American Physical Society},
  doi       = {10.1103/PhysRevLett.83.1467}
}

@phdthesis{ledoit1995essays,
  title  = {Essays on Risk and Return in the Stock Market},
  author = {Ledoit, Olivier Richard Henri},
  year   = {1995},
  school = {Massachusetts Institute of Technology}
}

@article{ledoit2003improved,
  title     = {Improved Estimation of the Covariance Matrix of Stock Returns with an Application to Portfolio Selection},
  author    = {Ledoit, Olivier and Wolf, Michael},
  journal   = {Journal of Empirical Finance},
  volume    = {10},
  pages     = {603--621},
  year      = {2003},
  publisher = {Elsevier}
}

@article{ledoit2004honey,
  title   = {Honey, I Shrunk the Sample Covariance Matrix},
  author  = {Ledoit, Olivier and Wolf, Michael},
  journal = {The Journal of Portfolio Management},
  volume  = {30},
  pages   = {110--119},
  year    = {2004}
}

@article{ledoit2004well,
  title     = {A Well-Conditioned Estimator for Large-Dimensional Covariance Matrices},
  author    = {Ledoit, Olivier and Wolf, Michael},
  journal   = {Journal of Multivariate Analysis},
  volume    = {88},
  pages     = {365--411},
  year      = {2004},
  publisher = {Elsevier}
}

@article{ledoit2017,
  author    = {Ledoit, Olivier and Wolf, Michael},
  title     = {Nonlinear Shrinkage of the Covariance Matrix for Portfolio Selection: {Markowitz} Meets {Goldilocks}},
  journal   = {The Review of Financial Studies},
  volume    = {30},
  pages     = {4349--4388},
  year      = {2017},
  month     = {06},
  issn      = {0893-9454},
  doi       = {10.1093/rfs/hhx052},
  url       = {https://doi.org/10.1093/rfs/hhx052},
  eprint    = {https://academic.oup.com/rfs/article-pdf/30/12/4349/24433779/hhx052.pdf}
}

@article{ledoit2022quadratic,
  title     = {Quadratic Shrinkage for Large Covariance Matrices},
  author    = {Ledoit, Olivier and Wolf, Michael},
  journal   = {Bernoulli},
  volume    = {28},
  pages     = {1519--1547},
  year      = {2022},
  publisher = {Bernoulli Society for Mathematical Statistics and Probability}
}

@book{de2020machine,
  title     = {Machine Learning for Asset Managers},
  author    = {Lopez de Prado, Marcos},
  year      = {2020},
  publisher = {Cambridge University Press}
}


\clearpage
\appendix
\renewcommand{\thesection}{Appendix~\Alph{section}}
\section{Spectral mapping and feasibility}
\label{sec:SpectralMapping}

\subsection{Correlation--PSD of class matrices \(C^{(\alpha)}\)}

\paragraph{Proof sketch.}
Each event atom \(A^{\pm}\) is positive semi-definite by construction: it is a rank-one
outer product with eigenvalues \(\{0,2\}\). For a given class \(\alpha\), the accumulator
\[
G_\alpha = \sum_{(i,j),\,e\in \alpha} w_e A^{\pm}(e), \qquad w_e > 0,
\]
is therefore a non-negative sum of PSD matrices and hence PSD itself.

To obtain \(C^{(\alpha)}\), we apply a congruence transform with the diagonal scaling
matrix \(D_\alpha = \mathrm{diag}(G_\alpha)\):
\[
C^{(\alpha)} = D_\alpha^{-\tfrac12} G_\alpha D_\alpha^{-\tfrac12}.
\]
Congruence transformations preserve positive semi-definiteness, so \(C^{(\alpha)}\) is PSD.
By construction \(C^{(\alpha)}\) is also symmetric with unit diagonal. Moreover, since \(G_\alpha\succeq 0\),
its entries satisfy the standard PSD (Cauchy--Schwarz / principal-minor) bound
\(|(G_\alpha)_{ij}|\le \sqrt{(G_\alpha)_{ii}(G_\alpha)_{jj}}\). It follows that the off-diagonal
entries of \(C^{(\alpha)}\) are bounded in \([-1,1]\). Thus each \(C^{(\alpha)}\) is correlation--PSD.

Finally, since the identity matrix \(I\) is PSD and convex combinations of PSD matrices are PSD,
the overall estimator \(S\) in \eqref{eq:impl-xi} is PSD whenever the collective squeeze does not exceed one.

\subsection{Exact PSD condition}

\paragraph{Spectral mapping under the neutral benchmark \(I\).}
Using the compact form in \eqref{eq:impl-xi}, we may write,
\[
S = (1-\xi)I + \xi P \;=\; I+\xi(P-I),
\]
where \(\xi=\sum_{\alpha\in\mathcal K}\eta_\alpha\) is the collective squeeze and \(P\) is the
corresponding normalized blend of class matrices.
Diagonalize \(P=U\Lambda U^\top\). Then
\[
S \;=\; (1-\xi)I + \xi U\Lambda U^\top
\;=\; U\bigl((1-\xi)I+\xi\Lambda\bigr)U^\top,
\]
so \(S\) and \(P\) share eigenvectors, with eigenvalues related by the affine map
\[
\lambda_i(S) \;=\; (1-\xi)\cdot 1 + \xi\,\lambda_i(P)
\;=\; 1+\xi\bigl(\lambda_i(P)-1\bigr).
\]
In particular,
\[
S\succeq 0
\iff
\lambda_{\min}(S)\ge 0
\iff
(1-\xi)+\xi\,\lambda_{\min}(P)\ge 0.
\]
Equivalently, for \(\xi>1\) (extrapolation), feasibility is
\[
\lambda_{\min}(P)\ \ge\ 1-\tfrac{1}{\xi},
\]
while for \(\xi\le 1\) (interpolation), PSD holds whenever the class matrices are PSD, since \(S\) is then a
convex combination of PSD matrices. This is the interpolation/extrapolation interpretation used in the main text.

\subsection{Basic IQ feasibility}

\paragraph{Basic IQ weights (\(\{\eta^2,\eta,1\}\), \(\eta\in[0,1]\)).}
Let \(\xi(\eta)=1+\eta+\eta^2\) and define
\[
P(\eta)=\frac{C^{(T)} + \eta\, C^{(B)}+ \eta^2\, C^{(W)}}{1+\eta+\eta^2}.
\]
Then \(\xi(\eta)\in[1,3]\), and for any \(\eta>0\) we are in extrapolation \(\bigl(\xi(\eta)>1\bigr)\).
Applying the exact PSD condition above with \(\xi=\xi(\eta)\) yields the feasibility threshold
\[
S(\eta)\succeq 0
\iff
(1-\xi(\eta))+\xi(\eta)\,\lambda_{\min}\!\bigl(P(\eta)\bigr)\ge 0
\iff
\lambda_{\min}\!\bigl(P(\eta)\bigr)\ \ge\ 1-\tfrac{1}{\xi(\eta)}
\ =\ \frac{\eta+\eta^2}{1+\eta+\eta^2}.
\]

\paragraph{Bounds in terms of class minima.}
Write \(m_k=\lambda_{\min}\!\bigl(C^{(k)}\bigr)\) for \(k\in\{T,B,W\}\).
Since \(\lambda_{\min}\) is concave on the PSD cone, for nonnegative weights summing to one,
\[
\lambda_{\min}\!\bigl(P(\eta)\bigr)
\;\ge\;
\frac{m_T+\eta m_B+\eta^2 m_W}{1+\eta+\eta^2}.
\]
A sufficient condition for feasibility is therefore
\[
\frac{m_T+\eta m_B+\eta^2 m_W}{1+\eta+\eta^2}
\ \ge\
\frac{\eta+\eta^2}{1+\eta+\eta^2},
\]
which rearranges to
\[
\eta^2(1-m_W)+\eta(1-m_B)\ \le\ m_T.
\]
For a necessary (coarse) bound, note that \(\lambda_{\min}\!\bigl(P(\eta)\bigr)\le m_{\max}\),
where \(m_{\max}:=\max\{m_T,m_B,m_W\}\). If \(S(\eta)\) is PSD, then in particular
\(\lambda_{\min}\!\bigl(P(\eta)\bigr)\ge (\eta+\eta^2)/(1+\eta+\eta^2)\), hence necessarily
\[
\frac{\eta+\eta^2}{1+\eta+\eta^2}\ \le\ m_{\max}
\quad\Longrightarrow\quad
\eta^2+\eta\ \le\ \frac{m_{\max}}{1-m_{\max}}.
\]
These bounds are the sufficient and necessary inequalities reported in the main text.

\subsection{Choosing \(\phi\) for \(\lambda_{\min}\) and \(\kappa\)}

For \(\widehat S(\phi)=(1-\phi)S+\phi I\), the extremal eigenvalues are affine in \(\phi\). If \(\alpha=\lambda_{\max}(S)\)
and \(\beta=\lambda_{\min}(S)\), then
\[
\lambda_{\max}(\widehat S(\phi)) = (1-\phi)\alpha + \phi,\qquad
\lambda_{\min}(\widehat S(\phi)) = (1-\phi)\beta + \phi.
\]
Solving \((1-\phi)\beta+\phi\ge\underline\lambda\) yields the expression in \eqref{eq:t-for-lmin}; solving
\[
\frac{(1-\phi)\alpha+\phi}{(1-\phi)\beta+\phi}\le K
\]
yields the expression in \eqref{eq:t-for-K}.

\section{Compact illustration}
\label{sec:CompactIllustration}

This section provides a small worked example to visualize the regimes and the role of the eigenfloor. The aim is not to benchmark but to make the mechanics concrete.

\paragraph{Setup.}
Take three assets \((n=3)\) and three channel classes \(B,W,T\).
In the full procedure the class matrices \(C^{(B)},C^{(W)},C^{(T)}\) are constructed from data by aggregating co-movement events through the squeezing template in Figure~\ref{fig:fig_1} and then normalizing once per class.
For this compact illustration we use the following correlation--PSD matrices, chosen to be numerically plausible and to make the regimes and the role of the eigenfloor transparent:
\[
C^{(B)}=
\begin{bmatrix}
1 & 0.30 & 0.20\\
0.30 & 1 & 0.25\\
0.20 & 0.25 & 1
\end{bmatrix},\quad
C^{(W)}=
\begin{bmatrix}
1 & -0.50 & -0.20\\
-0.50 & 1 & -0.30\\
-0.20 & -0.30 & 1
\end{bmatrix},\quad
C^{(T)}=
\begin{bmatrix}
1 & 0 & 0.80\\
0 & 1 & 0\\
0.80 & 0 & 1
\end{bmatrix}.
\]
In basic--IQ the class weights are \(\{1,\eta,\eta^2\}\) on \(\{T,B,W\}\) with \(\eta\in[0,1]\).
Hence the collective squeeze is \(\xi(\eta)=1+\eta+\eta^2\) and the blend is
\begin{equation}
\label{eq:P-eta-compact}
P(\eta)\;=\;\frac{C^{(T)}+\eta\,C^{(B)}+\eta^{2}C^{(W)}}{\,1+\eta+\eta^{2}\,},\qquad
S(\eta)\;=\;(1-\xi(\eta))I+\xi(\eta)P(\eta).
\end{equation}

\paragraph{Guaranteed vs.\ conditional regimes.}
Using the exact PSD condition from Appendix~\ref{sec:SpectralMapping}, \(S(\eta)\succeq 0\) if and only if
\((1-\xi(\eta))+\xi(\eta)\,\lambda_{\min}\!\big(P(\eta)\big)\ge 0\), that is,
\(\lambda_{\min}\!\big(P(\eta)\big)\ge 1-\tfrac{1}{\xi(\eta)}=\tfrac{\eta+\eta^2}{1+\eta+\eta^2}\) when \(\xi(\eta)>1\).
For \(\eta\in(0,1]\), \(\xi(\eta)>1\) and we are in extrapolation.
For three representative values (computed from the matrices above):

\medskip
\begin{center}
\begin{tabular}{@{}lcccc@{}}
\toprule
\(\eta\) & \(\xi(\eta)\) & \(\lambda_{\min}\!\big(P(\eta)\big)\) & threshold \(1-\tfrac{1}{\xi(\eta)}\) & PSD? \\
\midrule
\(0.0\) & \(1.00\) & \(0.200\) & \(0.000\) & yes \\
\(0.5\) & \(1.75\) & \(0.514\) & \(0.429\) & yes \\
\(1.0\) & \(3.00\) & \(0.729\) & \(0.667\) & yes \\
\bottomrule
\end{tabular}
\end{center}
\par\medskip

For these matrices the blend is sufficiently close to the identity that the conditional regime still yields PSD.

\paragraph{Effect on \(S(\eta)\) and the eigenfloor.}
At \(\eta=1\) the spectrum of \(S(\eta)\) is approximately
\[
\lambda\big(S(1)\big)\approx \{\,1.838,\ 0.977,\ 0.186\,\},
\]
so the matrix is PSD but moderately ill conditioned.
Two common stabilization goals can be met by a small identity reserve using \eqref{eq:eigfloor-def}--\eqref{eq:eigfloor-map}:

\begin{itemize}
\item \emph{Target condition number.} For \(K=5\), \eqref{eq:t-for-K} gives \(t\approx 0.185\), yielding
\[
\lambda\big(\widehat S(0.185)\big)\approx \{\,1.683,\ 0.981,\ 0.337\,\}.
\]
\item \emph{Target minimum eigenvalue.} For \(\underline{\lambda}=0.50\), \eqref{eq:t-for-lmin} gives \(t\approx 0.386\), yielding
\[
\lambda\big(\widehat S(0.386)\big)\approx \{\,1.514,\ 0.986,\ 0.500\,\}.
\]
\end{itemize}

Both choices move the spectrum toward \(1\) as predicted by \eqref{eq:eigfloor-map}, while preserving the squeezing representation via \eqref{eq:eigfloor-canonical} with rescaled class weights.

\paragraph{Remarks.}
(i) The example is intentionally small and uses fixed class matrices for clarity. In practice \(C^{(\alpha)}\) are constructed atomically from data and normalized once per class.  

(ii) When \(\xi\le 1\) (atomic--IQ) no floor is required; when \(\xi>1\) the eigenfloor provides a closed--form PSD repair and conditioning control while keeping the estimator inside the squeezing family.  

(iii) Larger asset panels and temporal weighting variants are reported in the accompanying empirical results (see Sections~\ref{sec:method}.\ref{sec:empericaldesign} \&~\ref{sec:method}.\ref{sec:empirical}).

\section{Sharpe ratio tests relative to atomic--IQ}
\label{sec:SharpeRatioTests}

This section describes the inference procedures used to compare the Sharpe ratios of all competing covariance estimators reported in the empirical results (Table~4) to AIQ1. Throughout, AIQ1 is treated as the benchmark. Our primary question is whether any alternative estimator delivers a higher Sharpe ratio than AIQ1. Accordingly, for each competitor \(k\) we formulate one-sided hypotheses of the form
\[
H_{0}: \text{SR}^{(k)} \ge \text{SR}^{(\text{AIQ1})}
\quad\text{vs.}\quad
H_{1}: \text{SR}^{(k)} < \text{SR}^{(\text{AIQ1})}.
\]
Rejection of \(H_{0}\) is interpreted as evidence that estimator \(k\) has a lower Sharpe ratio than AIQ1; failure to reject means that, based on Sharpe alone, \(k\) is at best comparable to AIQ1.

Because we test multiple competitors, the reported \(p\)-values should be interpreted as pairwise evidence relative to AIQ1. Unless stated otherwise, we do not apply a family-wise multiple-testing correction; the complete set of test statistics and confidence intervals is reported in the online supplement. To align with our study, in what follows assume \(T\equiv\tau=20\) and \(L=12\).

\subsection*{Setup and notation}

Let \(r_{t}^{(k)}\) denote the monthly excess return (net of the risk-free rate and after transaction costs) of the tangency portfolio constructed using estimator \(k\) in month \(t\), for \(t = 1,\dots,T\). We work with two Sharpe ratio objects:

\begin{enumerate}
\item The full-sample annualized Sharpe ratio,
\[
\widehat{\text{SR}}^{(k)} 
= \sqrt{12}\,\frac{\bar{r}^{(k)}}{\hat{\sigma}^{(k)}},
\qquad
\bar{r}^{(k)} = \frac{1}{T}\sum_{t=1}^{T} r_{t}^{(k)},\quad
\hat{\sigma}^{(k)} = \left(\frac{1}{T-1}\sum_{t=1}^{T} \bigl(r_{t}^{(k)} - \bar{r}^{(k)}\bigr)^{2}\right)^{1/2}.
\]

\item The rolling annualized Sharpe ratio computed over a window of \(W\) months (in the empirical work \(W=36\)). For each \(t \ge W\) we set
\[
\widehat{\text{SR}}_{t}^{(k)} 
= \sqrt{12}\,\frac{\hat{\mu}_{t}^{(k)}}{\hat{\sigma}_{t}^{(k)}},
\quad
\hat{\mu}_{t}^{(k)} = \frac{1}{W}\sum_{j=t-W+1}^{t} r_{j}^{(k)},\quad
\hat{\sigma}_{t}^{(k)} = \left(\frac{1}{W-1}\sum_{j=t-W+1}^{t} \bigl(r_{j}^{(k)} - \hat{\mu}_{t}^{(k)}\bigr)^{2}\right)^{1/2}.
\]
In addition to the standard monthly grid \((t = W,W+1,\dots,T)\), we also consider a thinned grid with a step of three months \((t = W,W+3,W+6,\dots)\) as a robustness check that reduces overlap across windows.
\end{enumerate}

For each competitor \(k\) we treat AIQ1 as the benchmark and define the Sharpe differences
\[
\Delta^{(k)} = \widehat{\text{SR}}^{(\text{AIQ1})} - \widehat{\text{SR}}^{(k)}
\quad\text{(full sample)},\qquad
D_{t}^{(k)} = \widehat{\text{SR}}_{t}^{(\text{AIQ1})} - \widehat{\text{SR}}_{t}^{(k)}
\quad\text{(rolling)}.
\]
A positive value of \(\Delta^{(k)}\) or \(D_{t}^{(k)}\) indicates that AIQ1 has the higher Sharpe ratio at the given horizon.

\subsection*{Full-sample Sharpe ratio tests}

We first test whether any estimator achieves a higher full-sample Sharpe ratio than AIQ1. For each competitor \(k\) we compute the observed Sharpe difference
\[
\widehat{\Delta}^{(k)} = \widehat{\text{SR}}^{(\text{AIQ1})} - \widehat{\text{SR}}^{(k)}.
\]
To obtain its sampling distribution under the null, we use a moving-block bootstrap over monthly excess returns. Let
\[
\mathbf{R}_{t}^{(k)} = \bigl(r_{t}^{(\text{AIQ1})},\,r_{t}^{(k)}\bigr)^{\top},
\qquad t=1,\dots,T,
\]
and stack these in a \(T\times 2\) matrix \(R^{(k)}\).

We form overlapping blocks of length \(L\) months,
\[
B_{1} = (\mathbf{R}_{1}^{(k)},\dots,\mathbf{R}_{L}^{(k)}),\;
B_{2} = (\mathbf{R}_{2}^{(k)},\dots,\mathbf{R}_{L+1}^{(k)}),\dots,\;
B_{T-L+1} = (\mathbf{R}_{T-L+1}^{(k)},\dots,\mathbf{R}_{T}^{(k)}),
\]
and then construct bootstrap samples by concatenating randomly selected blocks with replacement until at least \(T\) observations have been accumulated, truncating to length \(T\). For each bootstrap replication \(b=1,\dots,B\) we denote the resampled series by \(\{\mathbf{R}_{t}^{*(k,b)}\}_{t=1}^{T}\) and compute the corresponding Sharpe ratio difference
\[
\Delta^{*(k,b)} = \widehat{\text{SR}}^{*,(\text{AIQ1})} - \widehat{\text{SR}}^{*,(k)},
\]
based on the bootstrap excess returns. This yields an empirical distribution \(\{\Delta^{*(k,b)}\}_{b=1}^{B}\).

We report percentile \(95\%\) confidence intervals for \(\Delta^{(k)}\) as the \(2.5\)th and \(97.5\)th percentiles of this bootstrap distribution. For the one-sided test
\[
H_{0}: \text{SR}^{(k)} \ge \text{SR}^{(\text{AIQ1})}
\quad\text{vs.}\quad
H_{1}: \text{SR}^{(k)} < \text{SR}^{(\text{AIQ1})},
\]
the bootstrap \(p\)-value is approximated by
\[
p^{(k)}_{\text{full}} 
= \Pr\!\bigl(\Delta^{*(k)} \le 0\bigr)
\approx \frac{1}{B}\sum_{b=1}^{B} \mathbb{I}\{\Delta^{*(k,b)} \le 0\},
\]
that is, the fraction of bootstrap Sharpe differences that are less than or equal to zero. Small values of \(p^{(k)}_{\text{full}}\) provide evidence that \(\text{SR}^{(\text{AIQ1})}\) exceeds \(\text{SR}^{(k)}\). In the empirical results, GS2 and LS4 yield small one-sided \(p\)-values and bootstrap confidence intervals for \(\Delta^{(k)}\) that are bounded away from zero, indicating that their full-sample Sharpe ratios are significantly lower than AIQ1's. For the remaining estimators the confidence intervals include zero and the one-sided \(p\)-values are relatively large, so there is no statistical evidence that any alternative delivers a higher Sharpe ratio than AIQ1.

\subsection*{Rolling 36-month Sharpe ratio tests}

We next examine the relative Sharpe performance over rolling 36-month windows. For each competitor \(k\) and grid choice (monthly or every third month) we consider the sequence \(\{D_{t}^{(k)}\}_{t=1}^{T^{*}}\) and its sample mean
\[
\bar{D}^{(k)} = \frac{1}{T^{*}}\sum_{t=1}^{T^{*}} D_{t}^{(k)}.
\]
The parameter of interest is the unconditional mean
\[
\mu_{D}^{(k)} = \mathbb{E}[D_{t}^{(k)}],
\]
and we test the one-sided hypothesis
\[
H_{0}: \mu_{D}^{(k)} \le 0
\quad\text{vs.}\quad
H_{1}: \mu_{D}^{(k)} > 0.
\]
Under \(H_{0}\) estimator \(k\) is at least as good as AIQ1 in rolling Sharpe on average; rejection indicates that AIQ1 has the higher Sharpe ratio across 36-month windows.

Because the rolling windows overlap heavily, \(\{D_{t}^{(k)}\}\) is serially correlated and potentially heteroskedastic. We therefore use both Newey–West heteroskedasticity- and autocorrelation-consistent (HAC) standard errors and a moving-block bootstrap for the mean.

\paragraph{Newey–West inference.}  
Let \(u_{t}^{(k)} = D_{t}^{(k)} - \bar{D}^{(k)}\), and define the sample autocovariances
\[
\hat{\gamma}_{\ell}^{(k)} = \frac{1}{T^{*}}\sum_{t=\ell+1}^{T^{*}} u_{t}^{(k)} u_{t-\ell}^{(k)}, \qquad \ell = 0,1,\dots,q,
\]
for a truncation lag \(q\) measured in months. The Newey–West estimator of the long-run variance of \(T^{*1/2}\bar{D}^{(k)}\) is
\[
\widehat{\Omega}^{(k)}_{\text{NW}} 
= \hat{\gamma}_{0}^{(k)} + 2\sum_{\ell=1}^{q} w_{\ell}\hat{\gamma}_{\ell}^{(k)},
\qquad
w_{\ell} = 1 - \frac{\ell}{q+1}
\quad\text{(Bartlett weights)}.
\]
The HAC variance estimator for \(\bar{D}^{(k)}\) is then
\[
\widehat{\text{Var}}_{\text{NW}}(\bar{D}^{(k)}) = \frac{\widehat{\Omega}^{(k)}_{\text{NW}}}{T^{*}},
\]
with corresponding standard error \(\text{SE}_{\text{NW}}(\bar{D}^{(k)}) = \sqrt{\widehat{\text{Var}}_{\text{NW}}(\bar{D}^{(k)})}\) and \(t\)-statistic
\[
t^{(k)}_{\text{NW}} = \frac{\bar{D}^{(k)}}{\text{SE}_{\text{NW}}(\bar{D}^{(k)})}.
\]
For the one-sided test \(H_{0}:\mu_{D}^{(k)} \le 0\) vs \(H_{1}:\mu_{D}^{(k)} > 0\), the \(p\)-value is
\[
p^{(k)}_{\text{NW}} = 1 - \Phi\bigl(t^{(k)}_{\text{NW}}\bigr),
\]
where \(\Phi(\cdot)\) is the standard normal distribution function. Small values of \(p^{(k)}_{\text{NW}}\) indicate that the average 36-month Sharpe ratio of AIQ1 exceeds that of estimator \(k\).

\paragraph{Block bootstrap for the mean.}  
As a complementary robustness check, we also implement a moving-block bootstrap for the mean \(\bar{D}^{(k)}\), using the same block length \(L\) as in the full-sample analysis. We form overlapping blocks of \(\{D_{t}^{(k)}\}\) and generate bootstrap samples by concatenating random blocks with replacement until \(T^{*}\) observations are obtained. For each replication \(b\) we compute the bootstrap mean \(\bar{D}^{*(k,b)}\), yielding an empirical distribution \(\{\bar{D}^{*(k,b)}\}_{b=1}^{B}\). Percentile confidence intervals for \(\mu_{D}^{(k)}\) are obtained from the empirical quantiles of this distribution. A one-sided bootstrap \(p\)-value for \(H_{0}:\mu_{D}^{(k)} \le 0\) is approximated by
\[
p^{(k)}_{\text{boot}} 
= \Pr\bigl(\bar{D}^{*(k)} \le 0\bigr)
\approx \frac{1}{B}\sum_{b=1}^{B} \mathbb{I}\{\bar{D}^{*(k,b)} \le 0\}.
\]

\paragraph{Interpretation.}  
In the empirical results, the point estimates \(\bar{D}^{(k)}\) are positive for all competitors, indicating that AIQ1’s 36-month Sharpe ratio is higher on average than that of every alternative estimator. For GS2 and LS4, both the HAC \(t\)-statistics and the bootstrap confidence intervals support rejection of \(H_{0}\) on both the monthly and thinned grids, implying that their rolling Sharpe ratios are significantly lower than AIQ1’s. For LS1 and the nonlinear shrinkage estimators NLS6–NLS8, the Newey–West \(p\)-values on the monthly grid are also small, providing suggestive evidence of a Sharpe premium for AIQ1, although the bootstrap intervals are somewhat wider. For the remaining competitors we do not reject \(H_{0}\) at conventional levels, so there is no statistical evidence that any of them achieves a higher Sharpe ratio than AIQ1 over 36-month windows. Taken together with the full-sample tests, these results show that while several estimators are comparable to AIQ1 in Sharpe, none can be shown to outperform it, and the weaker methods are clearly dominated.

\end{document}